\documentclass[aps,prb,twocolumn,groupedaddress]{revtex4-1}
\usepackage{epsfig,amsmath,amssymb,graphics,color,calc}

\newcommand{\tr}{\mbox{Tr}}
\newcommand{\mus}{{\mu^{\star}}}
\newcommand{\mfs}{|f^{(2)}_\star|}

\let\a=\alpha \let\b=\beta  \let\d=\delta
   
\let\l=\lambda \let\m=\mu \let\n=\nu  
\let\s=\sigma   
   \let\G=\Gamma
 \let\Th=\Theta\let\L=\Lambda

\def\de{\mathrm d}

\def\to{\rightarrow}

\newcommand{\beq}{\begin{equation}} \newcommand{\eeq}{\end{equation}}

\allowdisplaybreaks

\begin{document}


\title{Nonperturbative fluctuations and metastability in a simple model: from observables to microscopic theory and back}


\author{C. Rulquin*, P. Urbani**, G. Biroli**, G. Tarjus* and M. Tarzia*}

\affiliation{*LPTMC, CNRS-UMR 7600, Universit\'e Pierre et Marie Curie,
bo\^ite 121, 4 Pl. Jussieu, 75252 Paris cedex 05, France\\**Institut de Physique Th\'eorique, Universit\'e Paris Saclay, CEA, CNRS, F-91191 Gif-sur-Yvette}



\begin{abstract}
Slow dynamics in glassy systems is often interpreted as due to thermally activated events between ``metastable'' states. This emphasizes the role of nonperturbative fluctuations, which is especially dramatic when these fluctuations destroy a putative phase transition predicted at the mean-field level. To gain insight into such hard problems, we consider the implementation of a generic back-and-forth process, between microscopic theory and observable behavior via effective theories, in a toy model that is simple enough to allow for a thorough investigation: the one-dimensional $\varphi^4$ theory at low temperature. We consider two ways of restricting the extent of the fluctuations, which both lead to a nonconvex effective potential (or free energy) : either through a finite-size system or by means of a running infrared cutoff within the nonperturbative Renormalization Group formalism. We discuss the physical insight one can get and the ways to treat strongly nonperturbative fluctuations in this context.
\end{abstract}


\maketitle

\section{Introduction}

Glass-forming liquids are systems whose salient physical properties appear controlled by ``nonperturbative'' phenomena. At least in the deeply supercooled regime, when approaching the glass transition, the dynamics is best described as activated and cooperative.\cite{berthier-biroli,tarjus-review,wolynes-book} The presence of such thermally activated processes is a prototypical example where the role of fluctuations must be treated in a  nonperturbative way. This is already well known in the case of standard homogeneous nucleation, for instance when a supersaturated vapor transforms into a stable liquid. The theoretical treatment of this problem involves rare, localized events, which are described as nucleation droplets in a phenomenological approach\cite{CNT} and  instantonic solutions of some free-energy functional at a field-theoretical level.\cite{langer}

Activation is connected to ``metastability''. In simple cases, the starting point of a theoretical description is a mean-field Landau free-energy functional, or a classical effective action in quantum field theory, which has several minima and is therefore nonconvex. This nonconvexity results from the absence of fluctuations in the mean-field description and introducing fluctuations in the theory leads to the exact free-energy or effective action with the needed convexity property. At the mean-field level, the deepest minimum is the stable thermodynamic state and the higher free-energy minima the metastable states (or false vacua in quantum field theory). The return to convexity has been theoretically described by explicitly accounting for excitations that are nonuniform in space, \textit{e.g.} spin waves for systems with a continuous symmetry and droplets for Ising-like models, which encode the nonperturbative nature of the phenomenon. In the Ising-like case, nucleation (and growth) of a droplet then describes the escape of the system from a metastable state to reach the stable state and leads to activated dynamics.\cite{langer}

In more involved situations encountered in systems undergoing a so-called fluctuation-induced first-order transition,\cite{coleman-weinberg,brazovskii} the mean-field theory does not provide a proper starting point as the relevant metastable states are themselves generated by fluctuations. One must therefore find a way to include fluctuations with wavelength up to some finite scale in order to produce metastability and then study the localized excitations of such an effective theory.\cite{strumia-tetradis,wetterichreview}

The difficulty in the case of glass-forming systems is even stronger. The nature of metastability and of the metastable states is much more elusive\cite{bouchaud04,franz05,dzero09} and the effective or coarse-grained landscape of minima and saddle-points is expected to be very complex, with a number of minima that is exponentially large in the system size. In this case, the ``bottom-up'' approach, deriving the behavior of macroscopic observables starting from the microscopic theory, \textit{i.e.}, interacting particles in the continuum, is just too difficult. Even computer simulations are of limited use for providing a full resolution because of the very fast growth of the equilibration time as one approaches the glass transition. 

A reasonable starting point would then be an effective theory that encodes the main physical ingredients while leaving out inessential ones. From a renormalization-group perspective, one would like to integrate out short length-scale fluctuations to obtain 
an effective theory that governs the long-range cooperative fluctuations. Numerical simulations
can be particularly useful in this respect since they take into account, virtually exactly but for systems of limited size, all fluctuations. 
Provided one has some intuition about the nature of the effective theory and of the associated local order parameter, one can then try to extract  the parameters of this theory from the exactly computed behavior of finite-size systems. This ``top-down'' approach would be instrumental in validating and establishing the proper effective theory.

Recently, there has been several numerical studies aimed at measuring the so-called Franz-Parisi potential $V(q)$\cite{franzparisi-potential} in models of glass-forming liquids.\cite{cammarotaV,berthierV,seoaneV,berthierjackV} This potential plays the role of a Landau 
free-energy where the order parameter is taken as the similarity or overlap between configurations. It is at the root of recent approaches that map the physics of supercooled liquids on effective theories of the random-field and random-bond Ising type.\cite{stevenson08,parisi-franz,jacquin12,franz13,CBTT14} Studying for an ensemble of reference liquid configurations the average value of $V(q)$ and its  fluctuations in finite-size systems should provide a way to access some of the parameters of the effective magnetic-like theory, as discussed in Ref. \onlinecite{CBTT14}.  Once the parameters are known, predictions of the effective theory on long length scales can be further checked against other numerical or experimental observations.

We think that this back-and-forth process, between microscopic theory and observable behavior via effective theories, is a key to solving the glass transition problem. It however requires a better understanding of the role played by nonperturbative fluctuations and the development of a theoretical approach able to capture them at all scales. The aim of this work is to study this problem on a toy model that is simple enough to be thoroughly investigated and use this as a benchmark for future work on glassy systems.

The model that we focus on is the one-dimensional $\varphi^4$ scalar field theory defined by the Hamiltonian:
\begin{equation}
\label{action}
H_L[\varphi]=\int_{0}^{L} \textrm{d}x \left[\frac c 2 (\partial_x \varphi)^2+V(\varphi(x))\right]
\end{equation}
where $L$ is the system size  and the local potential $V(\varphi)$ has a double-well form:
\begin{equation}
\label{bare_potential}
V(\varphi)=\frac r 4 (\varphi^2-1)^2
\end{equation}
We are interested in the low-temperature regime. There, the physics of the model is governed by strong nonperturbative fluctuations: these are kinks or domain walls between the positive and negative phases. These localized defects have a finite cost and their density is always finite, albeit very small at low temperature (it follows a Boltzmann distribution). However, by redistributing the positions of these kinks the system can gain entropy. It is therefore their presence that destroys the phase transition which is predicted at the mean-field level and remains in a perturbative treatment.

Even though the present analysis is motivated by the simplicity of the model, which allows a detailed study, this is more than an academic problem. The  one-dimensional $\varphi^4$ field theory is actually central to several fields in statistical physics and condensed-matter physics where it appears \textit{mutatis mutandis} in quite different problems: A Langevin dynamics in a double well,\cite{kurchanleshouches} quantum double wells\cite{coleman,TK00,Ao02,Za01,weyrauch06}, quantum-impurity problems\cite{andersonyuval,moore} are all different incarnations of this very same model (with sometimes extra-difficulties and decorations).

Let us now illustrate the main issues we are going to address in this work.

The first issue concerns what we called the ``top-down'' approach. Mirroring the current situation in glasses, the problem we consider is one in which we want to infer the parameters of a theory that we conjecture to be of the Ising/$\varphi^4$ type from the results of simulations and to further check that the effective microscopic theory we have in mind is the correct one. The input from simulations and other essentially exact computations that we consider as available knowledge is the probability of observing a given average value $\phi$ of the field in a system of finite size $L$, or, more precisely, its logarithm,
\begin{equation}
\label{finite-size_potential_definition}
U_L(\phi)=-\frac{1}{\beta L}\ln P_L(\phi) 
\end{equation} 
where $P_L(\phi) $ is the probability density to observe $\frac 1 L \int dx \varphi(x)$ equal to $\phi$ in a system of size $L$ with periodic boundary conditions and $\beta=1/(k_B T)$. We will call $U_L(\phi)$ {\it the finite-size effective potential} since it takes into account all fluctuations exactly, up to the length-scale $L$. In the ${L\rightarrow 0}$ limit it coincides with the bare potential $V(\phi)$, whereas in the thermodynamic limit it is equal to the free energy (exact effective potential) as a function of $\phi$. This function $U_L(\phi)$ is also known in quantum field theory as the ``constraint effective potential''.\cite{oraifeartaigh86}

Of course, in the case of the  one-dimensional $\varphi^4$ field theory we know from the start that the proper effective theory is just that  given in Eq. (\ref{action}). Nevertheless, the problem of inferring the bare parameters of the theory, $c,r$, and the energy of a kink from the behavior of finite-size systems is not straightforward. Moreover, understanding in detail the evolution of $U_L(\phi)$, \textit{i.e.}, how the change in shape of the finite-size effective potential is related to the progressive integration of nonperturbative fluctuations is also very instructive. The knowledge gained in the case of this simple problem will likely be useful for tackling more difficult and still unsolved ones.

The second issue is the development of a ``bottom-up'' approach that progressively takes into account fluctuations, including the nonperturbative ones, and allows one to eventually describe the macroscopic behavior. 
As explained before, numerical simulations are not helpful in this respect since by construction they can be performed on finite-size systems only. This applies more specifically to glassy systems where the time scales needed to relax large systems close to the glass transition are unreachable even with the best available computers. Extrapolations to obtain the thermodynamic limit are then often dangerous and quite unrealistic. Needless to say that this is of course not true for the one-dimensional theory studied here.  But as already stressed, the model is nonetheless used as a benchmark.

The theoretical method of choice for progressively bridging the gap from microscopic to macroscopic physics is the renormalization group (RG).\cite{wilson} Perturbative RG has been fully developed and understood since the 70's and 80's. The nonperturbative RG on the other hand has been the focus of an intense research only since the 90's: for reviews, see Refs. [\onlinecite{wetterichreview},\onlinecite{delamottereview}]. (At this point, we should acknowledge that there is always an ambiguity when using the adjectives ``perturbative'' and ``nonperturbative''. The former usually refers to an expansion in a few coupling constants and/or an expansion around the mean-field (gaussian) theory in powers of the difference between the spatial dimension and the upper critical dimension. The nonperturbative RG avoids such expansions and is based on quite different approximation schemes that can potentially describe strong-coupling physics and, a key point for us here, the effect of nonperturbative fluctuations.\cite{footnote}) 

\begin{figure}
 \includegraphics[width=0.45\textwidth]{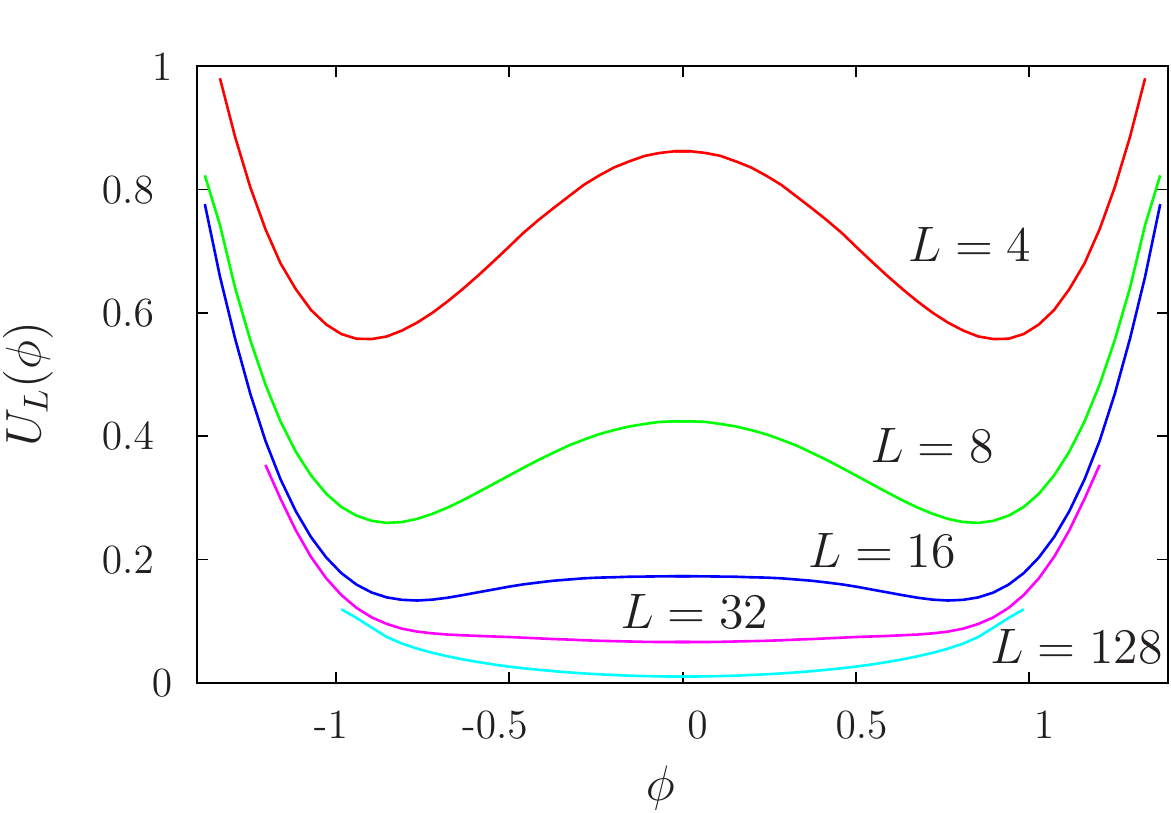}
 \label{fig:P(m)}
 \caption{Plot of $U_L(\phi)$ as a function of $\phi$ for different values of $L$, as obtained by MonteCarlo simulation:  $L=4,8,16,32,128$ for model parameters $r=2,c=4$.}
 \end{figure}

The nonperturbative (NP) RG method that is currently more used is the one introduced by Wetterich\cite{wetterich93,wetterichreview}. It has   been successfully applied to a variety of problems in high- and low-energy physics.\cite{wetterichreview,delamottereview} In the field of statistical physics it has led to nontrivial solutions of long-standing problems in frustrated magnets,\cite{delamotte-frustrated} disordered systems such as the random-field Ising model,\cite{gillesRFIM,tissierRFIM} and out-of-equilibrium dynamical phenomena.\cite{KPZ} The NPRG starts from an exact flow equation for the running effective action, $\Gamma_k[\phi]$, which is essentially the Legendre transform of the free-energy functional computed for an infinite system in which only fluctuations on length-scales less than $1/k$ have been integrated out. This running effective action at (length) scale $1/k$ coincides in the ultraviolet or microscopic limit, ${k\rightarrow \Lambda}$, with the bare Hamiltonian and in the infrared or macroscopic limit, $k\rightarrow 0$,  with the exact effective action (Gibbs free energy) as a function of the field $\phi(x)$. 

The success of this NPRG relies on determining a simple yet rich enough truncation of the exact NPRG equation. There are cases in which this strategy has been instrumental in tackling problems in which nonperturbative fluctuations are present: the $XY$ model in $2$ dimensions,\cite{wetterichXY} the random-field Ising model,\cite{gillesRFIM,tissierRFIM} or the return to convexity in the case of a first-order transition.\cite{wetterichreview,parola-reatto,tetradis92} However, the case we are considering in this work is somehow more difficult. There are strong nonlinear effects due to the interplay between the sharp changes in the field value on small length scales, which are associated with the kinks, and the long-wavelength field variations on a scale of the order of the distance between the kinks. We find that this one-dimensional physics, which one of course knows how to solve by other techniques, is harder to access via the NPRG and remains an unsolved problem within this approach. We will comment in conclusion about the (more favorable) situation in higher dimensions.

 \begin{figure}
 \includegraphics[angle=0,width=0.45\textwidth]{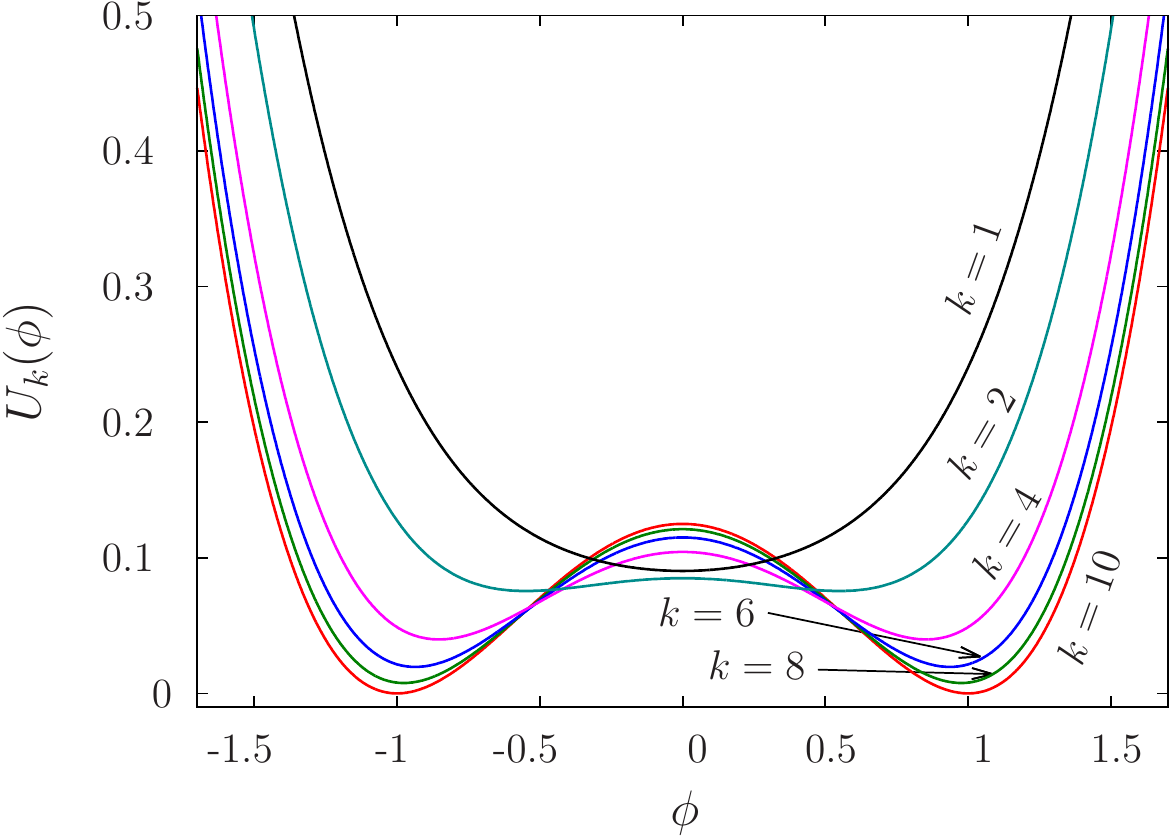}
 \caption{Plot of $U_k(\phi)=\Gamma_k(\phi)/(\beta L)$ as a function of $\phi$ for different values of $k$ obtained by using the Local Potential Approximation (LPA). The model parameters are $c=1$, $\beta=1$, and $r=1/2$. The flow equation is obtained with a regulator of the form\cite{litim01} $R_k(p)=(k^2-p^2)\Theta(k^2-p^2)$: $\partial_kU_k(\phi)=-\pi^{-1} U''_k(\phi)[U''_k(\phi)+k^2]^{-1}$. Note that there is a $k$-dependent but $\phi$-independent contribution that is not included (hence the apparent difference of behavior with that in Fig. \ref{fig:P(m)}).}
 \label{fig:Gamma(m)} 
 \end{figure}
 
In both of the above situations, \textit{i.e.}, either in a  system of finite size $L$ or within the NPRG in a system in the thermodynamic limit but in the presence of an infrared cutoff on fluctuations of wavelength larger than $1/k$, fluctuations are limited. As a result, the relevant potential, be it the finite-size one $U_L(\phi)$ or the running effective one $U_k(\phi)=\Gamma_k[\phi]/(\beta L)$, need not be convex. Just like in the mean-field limit where no fluctuations are taken into account, which in the present case leads to a Landau potential equal to the bare $V(\phi)$, metastability can thus be present. As the length scale over which fluctuations are allowed increases, \textit{i.e.}, with increasing $L$ or decreasing $k$, metastability should become less pronounced and in the macroscopic limit, $L\rightarrow \infty$ or $k \rightarrow 0$, both $U_L(\phi)$ and $U_k(\phi)$ should converge to the convex exact effective potential.

The typical evolution with $L$ of $U_L(\phi)$ is shown in Fig.~\ref{fig:P(m)} and that of the running effective potential $U_k(\phi)$ is plotted in Fig.~\ref{fig:Gamma(m)}. (In the latter case, we have obtained the result by using the so-called Local Potential Approximation (LPA)\cite{wetterichreview} of the exact NPRG equation.) The progressive disappearance of metastability is clearly observed in the two cases. The question we want to address in the former case is as follows: Say we are given some  numerical data in the form of Fig.~\ref{fig:P(m)}; how can one extract information about the corresponding effective theory and its parameters? On the other hand, in the latter case we would like to develop an approximation to the NPRG that is able to reproduce the main features associated with the nonperturbative fluctuations in the present model, for instance the known fact that the curvature of the potential in $\phi=0$ at low temperature is positive but very small as it behaves asymptotically as $\exp(-\beta S^\star)$, where $S^\star$ is the energy cost of a kink.
 
The rest of the paper is organized in two main sections, a first one where we address the top-down approach from finite-size studies and a second one where we discuss  the bottom-up one through the NPRG. To avoid disrupting the flow of the presentation, some technical details are relegated to appendices.

\section{The finite-size effective potential}
\label{sec:U_L}

In this section we study the behavior of the finite-size effective potential $U_L (\phi)$ and its evolution with the system size $L$. We first describe intuitively the shape of $U_L(\phi)$ and explain the ideas on how to extract the relevant quantities such as the correlation length $\xi$ and the surface tension $\gamma$ from the evolution of $U_L(\phi)$, by focusing in particular on the behavior of two quantities: the curvature $U_L(\phi)$ in $\phi=0$, $\kappa_L=U''_L(0)$, and the height of the barrier between the potential in $\phi=0$ and the minima in $\phi \simeq \pm 1$ (when present), $\Delta_L=U_L(0)-U_L(\pm 1)$. We then present detailed analytic results for $U_L (\phi)$ in the limit of zero temperature, which are obtained through the instanton technique, and use them as a benchmark to check the validity of our recipes for extracting the correlation length and the surface tension. Finally, we numerically determine the behavior of $U_L(\phi)$ at finite temperature, \textit{i.e.} at finite (but large) $\xi$. To this aim, we have combined Monte Carlo (MC) simulations and perturbation expansions based on a real-space RG and transfer matrix treatments. We then apply again the recipes for extracting the temperature dependence of $\xi$ and $\gamma$ and compare the output with the direct numerical computation of these quantities.

\subsection{The shape of $U_L (\phi)$ and its evolution with $L$}
\label{sec:shape_U_L}

At any given finite temperature, \textit{i.e.}, at any given finite correlation length $\xi$, if the system size goes to infinity, then the magnetization distribution goes to a Gaussian centered at $\phi=0$, due to the central limit theorem, and eventually converges to a Dirac delta function. (We use in this section the language of magnetic systems and call $\phi$ the magnetization, or, more properly, the magnetization density.) Thus, in the thermodynamic limit, the finite-size effective potential displays a unique (parabolic) minimum in $\phi=0$. On the other hand, at any given finite system size, as the temperature goes to zero and the correlation
length goes to infinity, the magnetization goes to either plus or minus one with probability one. For $\xi \gg L$ the finite-size effective potential is given by two symmetric minima centered in $\pm 1$. As a result, a nontrivial distribution of $P_L (\phi)$ and a nontrivial shape of $U_L(\phi)$ arise between the two opposite limits considered above.

\begin{figure}
{\includegraphics[scale=0.31]{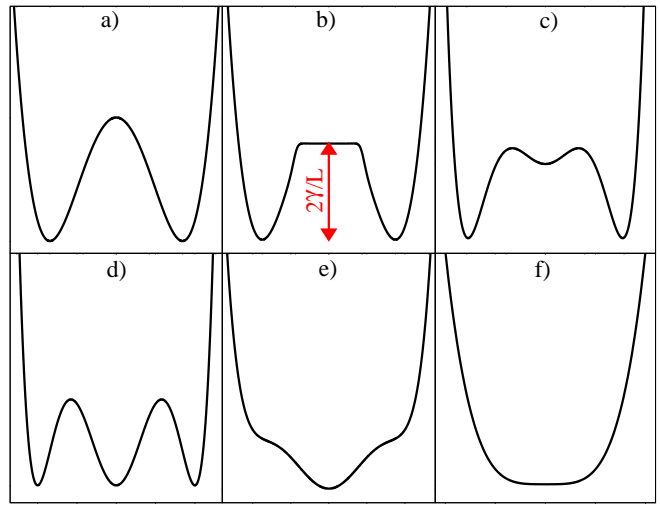}}
\caption{Sketch of the evolution of the shape of the finite-size effective potential $U_L(\phi)$ for different system sizes $L$:
a) $L < 2 \sigma$; b) $2 \sigma \ll L \ll \xi$; c) $L \lesssim \xi$; d) $L \simeq \xi$; e) $L \gtrsim \xi$; f) $L \gg \xi$.} 
\label{fig:UL}
\end{figure}

In order to figure out intuitively the evolution of $P_L(\phi)$ and $U_L(\phi)$ with the system size, let us first
focus on the typical configurations of the field $\varphi(x)$ which, at least at low enough temperature, dominate
the Gibbs measure. These are the configurations associated with the ground states of the system, corresponding to constant positive or negative magnetization profiles $\varphi(x) = \pm 1$, and the lowest excitations above them, involving domain walls (\textit{i.e.}, kinks and anti-kinks), which correspond to instantons that minimize the hamiltonian and connect positively 
and negatively magnetized regions. At low enough temperature (large enough correlation length), the typical
configurations of the field are thus well described by regions with almost constant $\pm 1$ magnetization separated by narrow domain walls. The width $\sigma$ of a domain wall is the typical size of an interface. The energy of a domain wall, $S^\star$, is, by definition, proportional to the microscopic surface tension $\gamma$ of the model, which is defined as the energy cost associated with the creation of an interface between two regions with opposite magnetization. It is easy to show (see below for more details) that the typical distance between domain walls is of the order of the correlation length $\xi$, which is proportional to $e^{\beta S^\star}$.

If $L$ is smaller than $2 \sigma$, no domain walls can be present in the system. (We consider periodic boundary conditions, so that the number of domain walls must be even.) Thus, $P_L(\phi) \simeq e^{- \beta L V (\phi)}$
and $U_L (\phi) \simeq V(\phi)$ (see Fig.~\ref{fig:UL}a). Therefore, from Eq.~(\ref{bare_potential}), $\Delta \simeq r/ 4$ and $\kappa \simeq - r$.

For $L> 2 \sigma$ but still much smaller than the correlation length $\xi$,  the probability of finding a domain wall is very small. The typical field configurations are then approximately constant $\pm 1$ magnetization profiles plus some small thermal fluctuations, whose amplitude depends on $V''(\phi=\pm 1) = 2 r$. On the other hand, configurations with zero magnetization correspond to field profiles with $2n$ domains walls, with $n \in \mathbb N^*$, that are suitably placed between $0$ and $L$. The thermodynamic weight of such configurations is proportional to $e^{- 2 n \beta S^\star}$ and the probability of having $\phi=0$ is obtained as
\begin{eqnarray} \label{eq:P0}
P_L (\phi=0) &\propto& \frac{L}{2} \Big [e^{- 2 \beta S^\star} + \frac{(L-4\sigma)^2}{8} \, e^{- 4 \beta S^\star} \\
\nonumber && + \frac{(L-6 \sigma)^4}{192} \, e^{- 6 \beta S^\star} + \ldots \Big ]\, ,
\end{eqnarray}
where the terms $(L-4 \sigma)^2/8$, $(L-6 \sigma)^4/192$, etc., correspond to the combinatorial factors accounting for the number of field configurations with $4$, $6$, etc., domain walls between $0$ and $L$ that have zero magnetization (see the next section
for more details). As long as $2\sigma \ll L \ll \xi$, all configurations with more than a single kink/anti-kink pair are  highly suppressed and their contribution can be neglected. Therefore, $P_L(\phi=0)$ is dominated by field profiles with only two domains walls. Since all such profiles have the same combinatorial factor (and thus the same probability), independently 
of the distance between the kink and the anti-kink, all intermediate magnetization values, sufficiently away from $\pm 1$, occur with approximately the same probability. As a result, for $2\sigma \ll L \ll \xi$ the finite-size effective potential $U_L (\phi)$ is given by two deep narrow symmetric minima around $\pm 1$ (whose curvature is simply given by $2 r/L$) that are  separated by a central region where $U_L(\phi)$ is approximatively constant. This is sketched in Fig.~\ref{fig:UL}b. The barrier height $\Delta_L = U_L(0) - U_L(\pm 1)$ is then given by $2 S^\star / L$, and the curvature in $\phi=0$ is $\kappa_L = U^{\prime \prime}_L (0) \simeq 0$.

The qualitative shape of $U_L(\phi)$ does not show any significant change until $L \lesssim \xi$. At this point, the terms of Eq.~(\ref{eq:P0}) corresponding to field configurations with more than two domain walls start to give a significant contribution to $P_L(\phi)$. Since there are exponentially many more configurations of the domain walls corresponding to zero magnetization with respect to configurations yielding positive or negative magnetization, $P_L (\phi)$ starts to develop a secondary maximum around $\phi=0$ as a result of this entropic effect.  Correspondingly, $U_L (\phi)$ develops a secondary minimum in zero, as sketched in Fig.~\ref{fig:UL}c. In this regime the behavior of the barrier height $\Delta_L$ and of the curvature $\kappa_L$ are model-dependent and cannot be
determined by simple intuitive argument: they must be computed in some explicit way, as we do in the following subsections.

For $L \simeq \xi$ the barrier $\Delta_L$ is expected to disappear as the value of the potential in $\phi=0$ crosses that in $\phi \simeq \pm 1$ (see Fig.~\ref{fig:UL}d). As $L$ further increases the minima in $\pm 1$ become higher and eventually disappear. However, the potential may still remain nonconvex, as illustrated in Fig.~\ref{fig:UL}e. Full convexity is recovered only for $L \to \infty$. It is then easy to show that the finite-size effective potential coincides with the Gibbs free-energy density (or exact effective potential) $U(\phi)$ of the system:
\begin{equation} \label{eq:Uinfinity}
U_L (\phi) = U(\phi) + o \left(\frac{1}{L} \right) \, ,
\end{equation}
where $U(\phi)$ is defined as the Legendre transform of the Helhmoltz free energy,
\begin{equation} \label{eq:gibbs}
U(\phi) = \beta^{-1}f(\beta,h) + h \phi \, ,
\end{equation}
where $h$ is the external magnetic field and $\langle \varphi \rangle = - \partial f(\beta,h) / \partial ( \beta h) = \phi$ (we have again used the magnetic language). As a consequence, for $L \gg \xi$ the finite-size effective potential is a convex function of $\phi$ and presents a unique minimum in $\phi=0$: see Fig.~\ref{fig:UL}f. In the thermodynamic limit the curvature $\kappa_L$ approaches $\kappa_{\infty} = U^{\prime \prime} (0)=\chi^{-1}$, where $\chi$ is the magnetic susceptibility defined as $\chi = \partial \langle \varphi \rangle / \partial ( \beta h) |_{h=0} = L ( \langle \varphi^2 \rangle - \langle \varphi \rangle^2)|_{h=0}$.

\begin{figure}
{\includegraphics[scale=0.32]{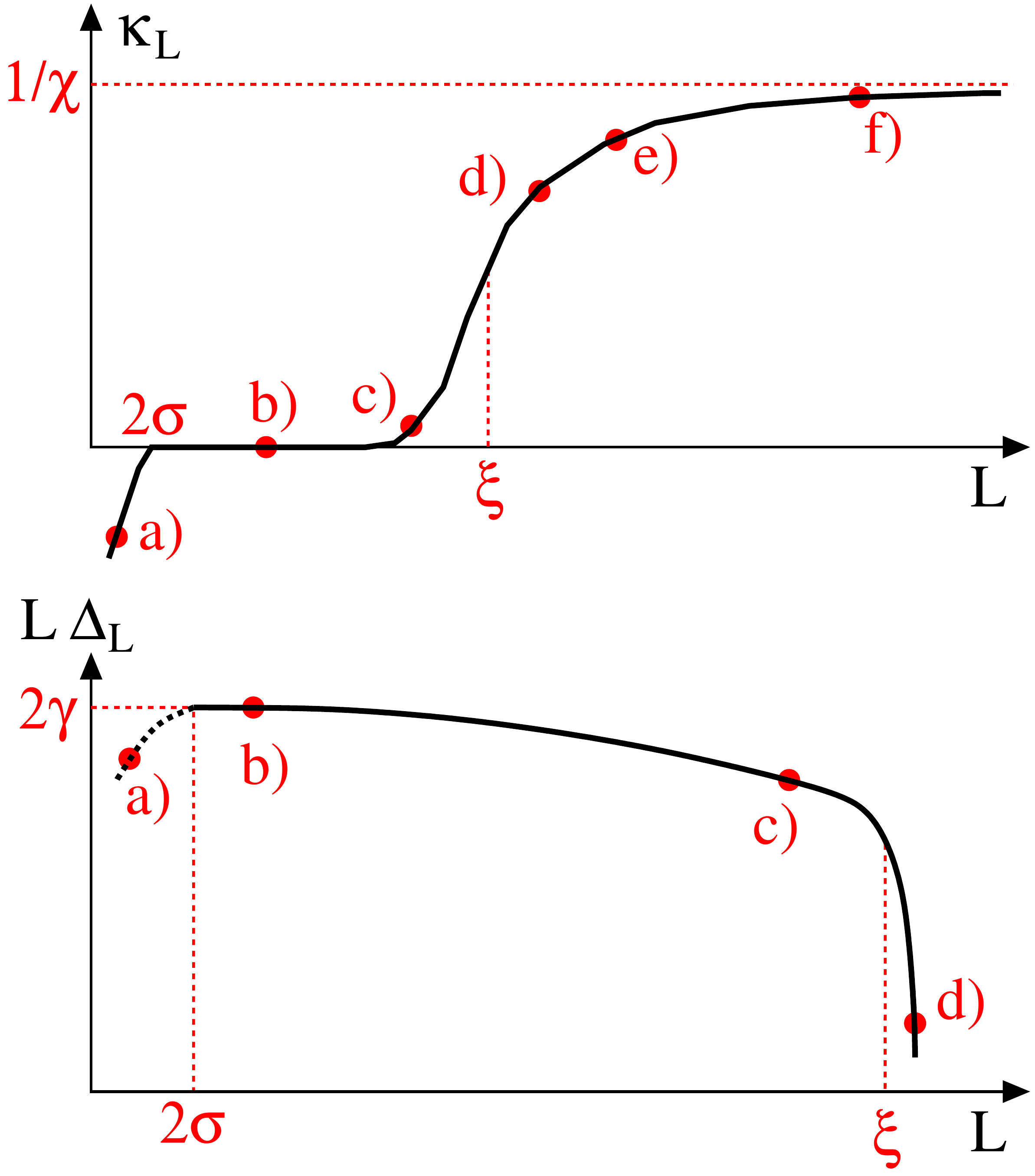}}
\caption{Schematic plot of the curvature $\kappa_L=U''_L(0)$ (top) and of the barrier height $\Delta_L=U_L(0)-U_L(\pm 1)$ times the system size $L$ (bottom) as a function of $L$. The barrier height is shown on a log-log plot. The labels a)-f) correspond to the shapes of $U_L(\phi)$ of Fig.~\ref{fig:UL}. Note that the behavior of $L \Delta_L$ at small $L\lesssim 2 \sigma$ is not universal and depends on the bare parameters.} 
\label{fig:schematic}
\end{figure}

Based on the intuitive arguments discussed above, we can qualitatively determine the behavior of the quantities of interest for us, $\kappa_L$ and $\Delta_L$, as a function of $L$.  They are schematically represented in Fig.~\ref{fig:schematic}. On very short length scales, $L < 2 \sigma$, the curvature is negative, $\kappa_L \simeq - r$. Then, for $2 \sigma \ll L \ll \xi$, $\kappa_L$ is approximately zero. For $L \lesssim \xi$, $\kappa_L$ starts to grow and for $L \to \infty$ it approaches $1/\chi$ as $1/L$. In turn, the barrier height $\Delta_L$ behaves roughly as $2 \gamma /L$ for $2 \sigma \ll L \ll \xi$ and rapidly vanishes for $L \gtrsim \xi$. 

We can therefore extract the important physical quantities by focusing on the behavior of $\kappa_L$ and $\Delta_L$. For instance, one possible recipe is to try to collapse the curves of $\kappa_L$ versus $L$ obtained at different temperatures onto a master curve by rescaling the $x$ and $y$ axes by adjustable parameters. The parameters that provide the best collapse should then be $\chi(T)^{-1}$ for $\kappa_L$ ($y$ axis) and $\xi(T)$ for $L$ ($x$ axis). Another possibility would be to plot $L \Delta_L$ as a function of $L$ and, knowing the correlation length $\xi(T)$ from the previous operation, to look for a plateau or a region of weak dependence on $L$ for $L<\xi$: at low enough temperature, the height of the plateau should then be twice the surface tension $\gamma(T)$. (Alternatively, one could do a log-log plot as in Fig. \ref{fig:schematic}b.) In the next  subsections we will implement and check these ideas in a quantitative way. 

\subsection{$U_L(\phi)$ in the $T \to 0$ ($\xi \to \infty$) limit}
\label{sub:finite-size_instantons}

As explained above, at very small temperature the Gibbs measure is dominated by the ground state of the system and the lowest excitations above it. The ground states of Eq.~(\ref{action}) correspond to constant field configurations $\varphi(x) = \pm 1$.
The lowest excitations above the ground states correspond to nonuniform kink and anti-kink profiles that are obtained by minimizing the Hamiltonian:
\begin{equation}
\left . \dfrac{\delta H_L [\varphi(x)]}{\delta \varphi(x)} \right |_{\varphi^\star} = 0 
\,\,\,\,\, \Rightarrow \,\,\,\,\,
c \, \dfrac{\partial^2 \varphi^\star(x)}{\partial x^2} = \left .\dfrac{\partial V(\varphi(x))}{\partial \varphi(x)} \right|_{\varphi^\star} \, ,
\end{equation}
with the boundary conditions $\varphi^\star (x \to - \infty) = \mp 1$ and $\varphi^\star (x \to + \infty) = \pm 1$.
This differential equation can be solved exactly for the $\varphi^4$ theory in $d=1$, yielding
\begin{equation} \label{eq:phistar}
\varphi^\star(x)= \pm \tanh(x/\sigma) \, ,
\end{equation}
with $\sigma = \sqrt{2c / r}$. 
The energy cost associated to such domain walls can be obtained by plugging Eq.~(\ref{eq:phistar}) into Eq.~(\ref{action}), which gives $S^\star = \int H_L [\varphi^\star(x)] \, \textrm{d} x \approx \sqrt{8 r c / 9}$. Note that we can make more precise now the notion of  ``low-temperature regime'': It is obtained by letting either $\beta$ or $r$ be large, such that the Boltzmann factor associated with the presence of a domain wall, $\exp(-\beta \sqrt{8rc/9})$, is much smaller than 1.

Small fluctuations of the field around the instantonic profile can be easily taken into account at a Gaussian level. After expanding the Hamiltonian around the instantonic solution up to second order in $\delta \varphi (x)= \varphi(x) - \varphi^\star (x)$, the thermodynamic weight of a single instanton is expressed as
\begin{equation}
Z_1 \simeq e^{- \beta S^\star} \int \mathcal{D} \varphi \, 
e^{- \frac{1}{2}  \int \textrm{d} x \, \textrm{d} y \, \left . \frac{\delta^2 H_L}{\delta \varphi (x) \delta \varphi(y)}\right|_{\varphi^\star} \delta \varphi(x) \, \delta \varphi(y) } \, .
\end{equation}
In order to compute the functional integral above one thus need to diagonalize the operator corresponding to the kernel
\begin{equation}
M(x,y) = \left[ - c\frac{\partial^2}{\partial x^2} + V^{\prime \prime} (\varphi^\star (x)) \right] \delta (x - y) \, ,
\end{equation}
which yields
\begin{equation}
Z_1 \simeq e^{- \beta \tilde{S}^\star} = e^{- \beta S^\star + \frac{1}{2} \ln \left( 2 \pi/\textrm{det} M \right) } \, .
\end{equation}
In the present low-temperature limit, it can be shown that $\textrm{det} \, M \sim \beta r$ and, thus, $\tilde{S}^\star \simeq S^\star - (1/2)\beta^{-1} \ln (\beta r) + \textrm{O}(\beta^{-1})$.

At very low temperature the typical configurations of the field are therefore described by a dilute gas of domain walls separated by regions with constant $\varphi = \pm 1$. The partition function of the system can thus be written as a sum over the number $n$ of kink/anti-kink pairs (as discussed above, the number of domain walls must be even to be compatible with the periodic boundary conditions) weighted by the energy cost $e^{-2 k \beta \tilde{S}^\star}$ times an appropriate combinatorial coefficient $I_{2n}$ accounting for all the possible configurations of the positions of $2n$ domain walls between $0$ and $L$:
\begin{equation}
Z_L = \sum_{n=0}^{[L/2\sigma]} I_{2n}(L) \,e^{-2 n \beta \tilde{S}^\star} \, ,
\label{zed}
\end{equation}
where $[x]$ denotes the integer part of $x$.
Note that, since the instantons have a finite width $\sigma$, we cannot place more than $(L/2\sigma)$ kink/anti-kink pairs between $0$ and $L$. The problem of determining the combinatorial coefficients $I_{2n}$ is equivalent to computing the entropy of a gas of $2n$ hard spheres of size $\sigma$ on a ring of length $L$ (see Appendix~\ref{app:instantons}). The resulting expression is
\begin{equation}
\label{eq_I_2n}
I_{2n}(L)=\dfrac{1}{n}\,\dfrac{L}{(2n-1)!} (L-2n\sigma)^{2n-1} \, .
\end{equation}
Two length scales thus naturally emerge from the calculation:  $\sigma$, the typical size of an interface, and $e^{\beta \tilde{S}^\star}$, which corresponds to the typical distance between two consecutive instantons and can be shown to be (twice) the correlation length $\xi$ of the system.

After introducing the rescaled variables $\zeta = L/e^{\beta \tilde{S}^\star}$ and $\alpha = \sigma/L$, the partition function finally reads
\begin{equation} 
\label{eq:ZL-instantons}
Z_L(\zeta,\alpha) = 2 \sum_{n=0}^{[1/(2\alpha)]} \, \dfrac{\zeta^{2n}}{(2n)!} (1-2n\alpha)^{2n-1} \, .
\end{equation}

The computation of the magnetization probability distribution $P_L(\phi)$ in the $T \to 0$ limit can be carried out in a similar way. Note that an analogous computation has already been done for the Ising model in $d=1$~\cite{antal} (see also below).

For each given instantonic configuration with $2n$ alternate kinks and anti-kinks we define $x_i$, $i=1, \ldots, 2n$, as the lengths of the regions with constant $\varphi = \pm 1$. In terms of these variables, the extensive magnetization $M$ reads
\begin{equation} \label{eq:M-instantons}
M = \int_0^L \varphi(x) \, \textrm{d} x = \pm \sum_{i=1}^n \left( x_{2i - 1} - x_{2i} \right) \, .
\end{equation}
Note that thanks to the translational invariance, one can choose without loss of generality to place the first domain wall at $x=0$. 
The sign of $M$ in front of the sum thus depends on whether the the first instanton is from $\varphi = -1$ to $\varphi = +1$ or \textit{vice versa}. 
Since each domain wall has a width $\sigma$, we also have that
\begin{equation} \label{eq:L-instantons}
\sum_{i=1}^{2n} x_i = L - 2n \sigma \, .
\end{equation}
In consequence, the extensive magnetization is bounded as $|M| \le L - 2 n \sigma$. When enforcing the constraints given by Eqs.~(\ref{eq:M-instantons}) and (\ref{eq:L-instantons}) one obtains
\begin{eqnarray}
P_L(M) &=& \frac{1}{Z_L} \Big[ \delta(M-L) + \delta(M+L) \\
\nonumber
&& + 2 \!\!\!\!\!\!\!\! \sum_{n=1}^{[(L-|M|)/2\sigma]} \! J_{2n}(M,L) \, e^{- 2n \beta \tilde{S}^\star} \Big] \, ,
\end{eqnarray}
where $Z_L$ is defined in Eq.~(\ref{zed}). Again, the combinatorial factors $J_{2n}(M,L)$ can be computed exactly (see Appendix~\ref{app:instantons}). After introducing the rescaled variables $\zeta$, $\alpha$ defined above and the magnetization density $\phi=M/L$ and using the fact that $\delta(L\phi) =(1/ L) \delta(\phi)$ and $P_L(L\phi) =(1/L) P_L(\phi)$, we finally obtain:
\begin{eqnarray} 
\label{eq:PLm-instantons}
P_L(\phi) &=& \dfrac{1}{Z_L(\zeta,\alpha)} \bigg[ \delta(\phi-1) + \delta(\phi+1) \\
\nonumber
&& + \, 2 \!\! \sum_{n=1}^{[(1-|\phi|)/2\alpha]} \! (\zeta/2)^{2n} \,\dfrac{\left[(1 - 2n\alpha)^2 - \phi^2\right]^{n-1}}{n!(n-1)!} \bigg] \, ,
\end{eqnarray}
where $Z_L(\zeta,\alpha)$ is given in Eq.~(\ref{eq:ZL-instantons}). It is easily checked that $P_L(\phi)$ is properly normalized, $\int_{-1}^{1} P_L (\phi) \, \textrm{d} \phi = 1$. One also finds that in the limit $\sigma \to 0$, \textit{i.e.}, when the domain walls become infinitely sharp, and for $\tilde{S}^\star = 2 J$, Eqs.~(\ref{eq:ZL-instantons}) and (\ref{eq:PLm-instantons}) give back the exact results derived for the one-dimensional Ising model.\cite{antal}. These calculations are explicitly done in Appendix~\ref{app:instantons}.

The finite size effective potential and its evolution with the system size can be now explicitly determined in the $T \to 0$ limit from the relation in Eq.~(\ref{finite-size_potential_definition}). $U_L(\phi)$ behaves as anticipated in the previous section: It presents two narrow minima in $\phi = \pm 1$, corresponding to the $\delta$-functions,\cite{width} and a secondary minimum in $\phi=0$ due to the entropic term in Eq.~(\ref{eq:PLm-instantons}). As $L$ increases (\textit{i.e.}, $\alpha$ decreases) the minimum in $\phi=0$ becomes deeper and deeper, as the sum over $n$ in Eq.~(\ref{eq:PLm-instantons}) grows exponentially fast with $\zeta/\alpha$. For $L \simeq \xi$ the value of the minimum in $\phi=0$ crosses that of the two symmetric minima in $\phi=\pm 1$ (strictly speaking the value of the minima in $\phi=\pm 1$ is defined only for a nonzero temperature;\cite{width} otherwise, one has to consider the weight of the delta peaks). Nevertheless, at any finite $L$ the potential remains  nonconvex due to the vestiges of the two minima in $\pm 1$. It is only in the thermodynamic limit that $U_{L \to \infty} (\phi)=U(\phi)$ recovers full convexity.

\begin{figure}
{\includegraphics[scale=0.65]{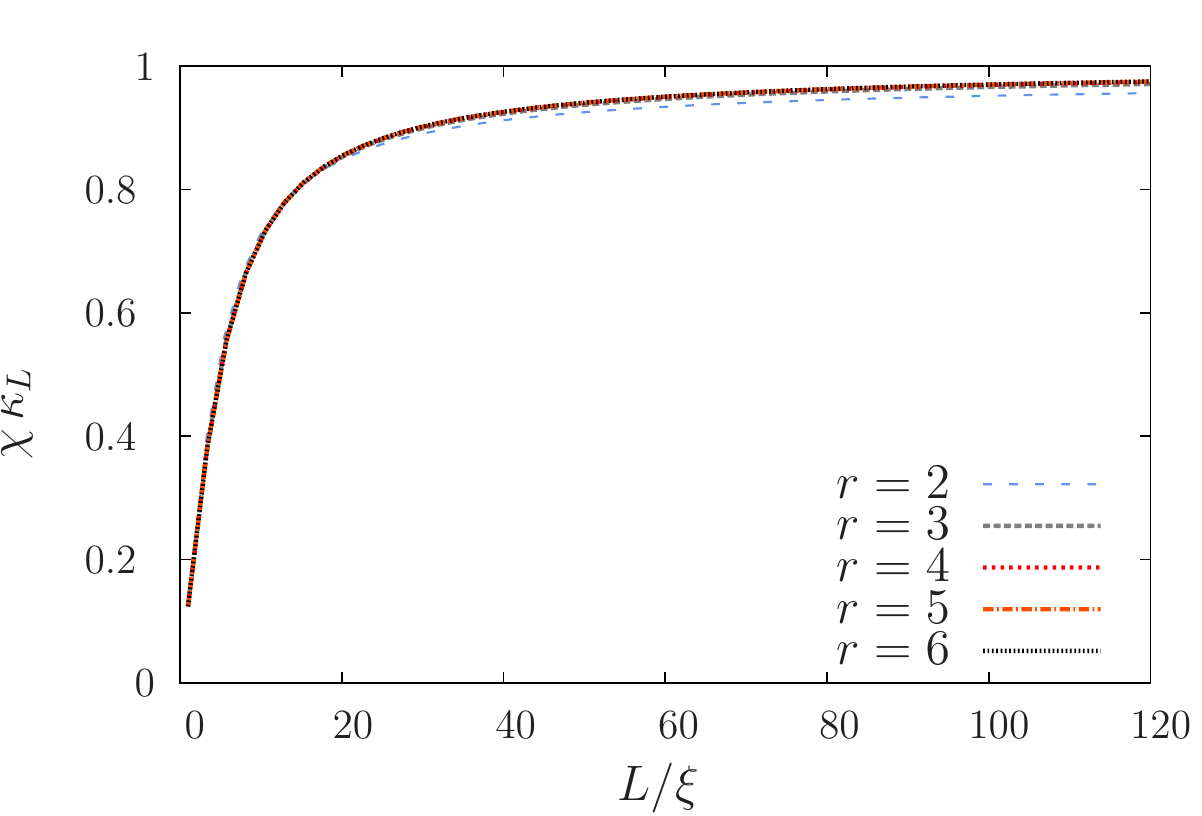}}
\caption{Rescaled curvature $\chi \kappa_L$ as a function of $L/\xi$ for the $\varphi^4$ field theory in $d=1$ in the low-temperature regime for $\beta=1$, $r$ varying from $2$ to $6$, and for $c=2 r$.} 
\label{fig:kappa-instantons}
\end{figure}

\begin{figure}
{\includegraphics[scale=0.65]{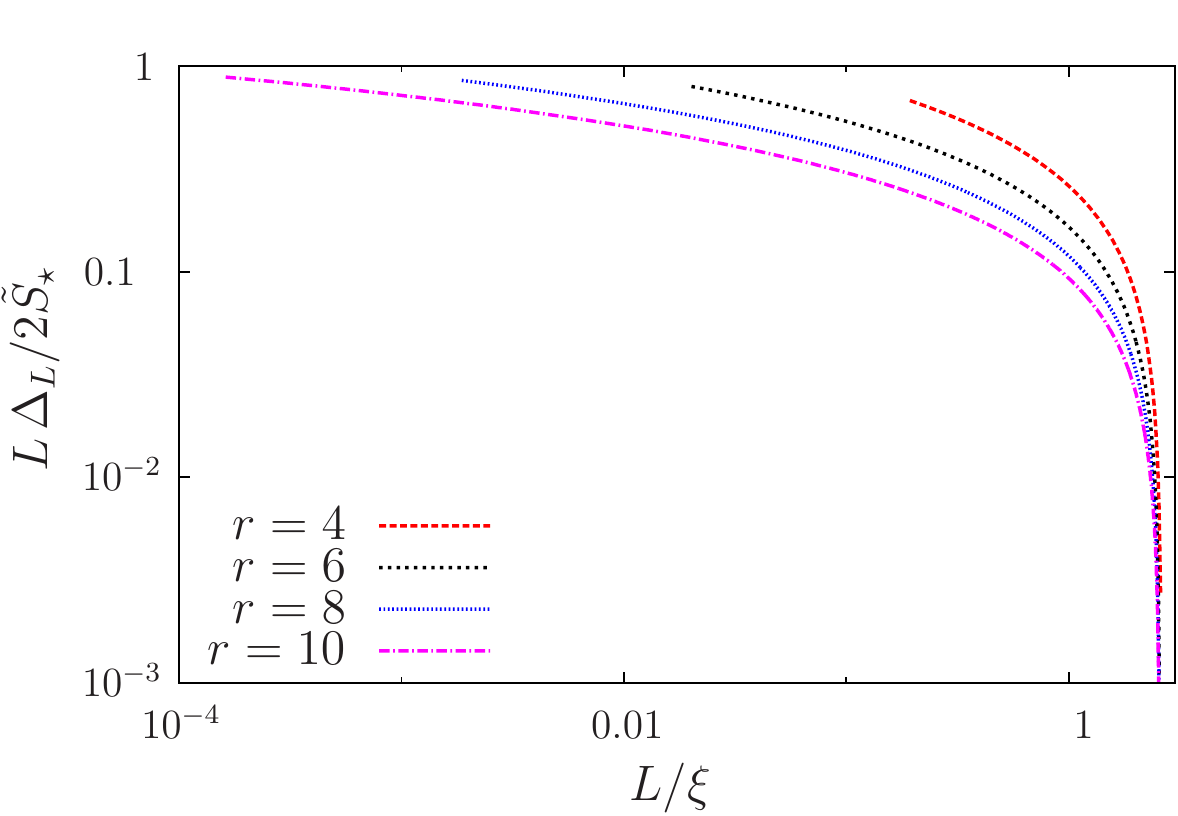}}
\caption{Log-log plot of the rescaled barrier, $L \Delta_L / (2 \tilde{S}^\star)$, as a function of $L/\xi$ 
for the $\varphi^4$ field theory in $d=1$ for $\beta=1$, $r$ varying from $4$ to $10$, and $c=2 r$. The curves stop on the low side for $L=2\sigma +a$ and the barrier is not defined within the instanton treatment for $L<2\sigma$.}
\label{fig:delta-instantons}
\end{figure}

From Eqs.~(\ref{finite-size_potential_definition}), (\ref{eq:ZL-instantons}), and (\ref{eq:PLm-instantons}), one can compute all the desired  characteristics of $U_L(\phi)$, such as the curvature in $\phi=0$, $\kappa_L$, and the barrier height (when present), $\Delta_L$.
Following the ideas presented in the previous section, we plot in Fig.~\ref{fig:kappa-instantons} the curvature $\kappa_L$ multiplied by the magnetic susceptibility $\chi$ as a function of the system size $L$ divided by the correlation length $\xi$, for different temperatures.\cite{chi} We have set $\beta=1$, $c = 2 r$ and a range of $r$ from $2$ to $6$ (as discussed above, the low-temperature limit here means that $\exp(-\beta \sqrt{8rc/9})=\exp(-4r/3) \ll 1$). The curves show a perfect collapse, as expected. Via the instanton calculation we indeed have access to all physical quantities of the present simple model. This is a consistency check of the recipe discussed above and an illustration of the range of temperatures where the asymptotic results apply. We show in Fig.~\ref{fig:delta-instantons} the evolution of the barrier height $\Delta_L$ multiplied by the system size $L$ and divided by twice the domain-wall (free) energy $\tilde{S}^\star$ as a function of $L/\xi$ for $\beta=1$, $c=2r$, and for $r$ varying from $4$ to $10$. This log-log plot is very similar to the sketch in Fig. \ref{fig:schematic}b. The value $L \Delta_L / (2 \tilde{S}^\star) \approx 1$ is observed for $L \sim 2 \sigma$, which corresponds to very small values of $\xi/L$, especially at low temperature. There is then a broad regime, up to $\xi/L \lesssim 1$,  where one observes a small decay, by less than a factor of $10$. Finally, for $L \gtrsim \xi$ there is a fast (exponential) decay. 

These plots validate at a quantitative level the proposed ways of extracting the parameters of the theory from the behavior of the finite-size effective potential. We now turn to the same exercise but in the finite temperature regime where the analytical solution via the instantons is no longer a sufficient description.

\subsection{$U_L(\phi)/L$ for finite but large $\xi$}

In this section we apply and test the empirical recipes to extract $\xi$, $\chi$ and $\gamma$ proposed above on a system at large but finite correlation length (corresponding to small but finite temperature). We obtain a numerical estimate of the finite-size effective potential $U_L(\phi)$ for a range of values of $L$ through several methods.

First, we have performed MC simulations. To this aim, we first discretize the continuum field theory of Eq.~(\ref{action}) by replacing the gradient by its discrete lattice version. The Hamiltonian thus becomes
\begin{equation} \label{action-discr}
H_L(\{\varphi_i\}) = a\sum_{i=1}^L \left[ \frac{c}{2 a^2} \left( \varphi_i - \varphi_{i+1} \right)^2 + V(\varphi_i) \right] \, .
\end{equation}
We set the lattice spacing $a$ to $1$ (note that for $c = 2 r$ the width of a domain wall is then $\sigma = 2 > a$) and we consider periodic boundary conditions: $\varphi_{L+1} = \varphi_1$. 

The numerical simulations are performed with a Metropolis algorithm: At each time step we pick a site $i$ at random, and attempt to change the value of $\varphi_i$ by a random quantity, $\delta \varphi_i$, extracted from a gaussian distribution with zero mean and variance $\sigma_\varphi$. We then compute the energy difference $\Delta H_L =c \delta \varphi_i (  \delta \varphi_i + 2 \varphi_i - \varphi_{i+1} - \varphi_i ) + V(\varphi_i + \delta \varphi_i) - V(\varphi_i)$ and accept the move with the Metropolis probability $p = \min \{1, e^{-\beta \Delta H_L} \}$. Time is advanced by $1/L$. The typical width of the field shifts $\sigma_\varphi$ is optimized recursively during the dynamics by enforcing that the acceptance rate of the moves (averaged over the last $100$ MC steps) is approximately equal to $0.3$.

We start from a given initial condition (for instance $\varphi_i = +1 \; \forall i$) and let the system evolve and equilibrate. The equilibration time $\tau$, which of course depends on $\beta$, $r$, and $L$, can be extracted from the exponential decay of dynamical correlation functions such as $(1/L) \sum_i \langle \varphi_i (t + t^\prime) \varphi_i( t^\prime) \rangle \simeq \langle \varphi^2 \rangle \, e^{-t/\tau}$.
In order to compute the magnetization probability distribution, $P_L(\phi)$, we measure the instantaneous magnetization $\varphi(t) = (1/L) \sum_i \varphi_i(t)$ at regular time intervals corresponding to several times the equilibration time, say $10 \tau$. This allows us to make sure that the values of $\varphi(t)$ measured during the dynamics are statistically independent. In this way we construct a histogram of the magnetizations which gives an estimate of $P_L(\phi)$ and, from Eq.~(\ref{finite-size_potential_definition}), we obtain $U_L (\phi)$. Results for $\beta=1$, $r=2$, $c=2r$, and $L$ varying from $4$ to $128$ are shown in Fig.~\ref{fig:P(m)}.

Note that in order to obtain an accurate enough estimate of $P_L(\phi)$ and of $U_L (\phi)$, we need to sample rare events, which take place with an exponentially small probability in the system size. As a consequence, the number of measurements of the instantaneous magnetization must scale exponentially with $L$. Since the computational time of a single MC step scales linearly with the system size, 
this implies that the total computational time of our MC simulations scales as $\tau L e^{L}$. Therefore, MC results are limited to not too large values of $L$, typically $L\lesssim 10^2$.

In order to overcome this limitation and study larger system sizes, we have used a $1/L$ perturbation expansion combined with an exact computation of the (Helmholtz) free-energy of the model through both a real-space RG approach and a transfer-matrix technique. 

Let us start with the definition of the magnetization probability distribution,
\begin{equation}
P_L(\phi) = \frac{\tr_{\{\varphi_i\}} \delta(L\phi - \sum_i \varphi_i) e^{- \beta H_L}}{\tr_{\{\varphi_i\}} e^{-\beta H_L}} \, ,
\end{equation}
where $\tr_{\{\varphi_i\}} \equiv \int \prod_i \textrm{d} \varphi_i$. By using the integral representation of the $\delta$-function, one easily obtains
\begin{equation} 
\label{eq:PLm-saddle}
P_L(\phi) = e^{L \beta f_L (\beta,0)} \int_{-i \infty}^{i \infty} \textrm{d}\mu \, e^{-L[\beta f_L (\beta, \mu) + \mu \phi]} \, ,
\end{equation}
and
\begin{equation} \label{eq:ULm-saddle}
U_L(\phi) = - \frac{1}{\beta L} \ln \int_{-i \infty}^{i \infty} \textrm{d}\mu \, e^{-L[\beta f_L (\beta, \mu) + \mu \phi]} 
- f_L (\beta,0) \, ,
\end{equation}
where $f_L (\beta,\mu)$ is the Helmholtz free-energy density of a system of size $L$ in the presence of an external uniform magnetic field $\mu/\beta$:
\begin{equation}
f_L (\beta,\mu) = - \frac{1}{\beta L} \ln \tr_{\{\varphi_i\}} e^{- \beta H_L + \mu \sum_i \varphi_i} \, .
\end{equation}
For large enough $L$ the integral in Eqs.~(\ref{eq:PLm-saddle}) and (\ref{eq:ULm-saddle}) is dominated
by the maximum in $\mu = \mus$, which is given by\cite{footnote_contour}
\begin{equation}
\left . \frac{\partial \beta f_L (\beta,\mu)}{\partial \mu} \right |_{\mus} + \phi = 0 \, .
\end{equation}
Expanding the argument of the exponential around $\mus$ leads to
\begin{eqnarray} \label{eq:exp}
\nonumber
\beta f_L (\beta,\mu) + \mu \phi & = & \beta f_L(\beta,\mus) + \mus \phi + \frac{\beta}{2} f^{(2)}_\star (\delta \mu)^2 \\
\nonumber
&& \,\,\,\,\, + \frac{\beta}{3!} f^{(3)}_\star (\delta \mu)^3 + \frac{\beta}{4!} f^{(4)}_\star (\delta \mu)^4 + \ldots
\, ,
\end{eqnarray}
where $f^{(n)}_\star = \partial^n f_L (\beta,\mu) / \partial \mu^n|_{\mus}$ and $\delta \mu = \mu - \mus$. One can thus treat all terms beyond the gaussian level in a perturbative way and obtain a systematic expansion of $P_L(\phi)$ and $U_L(\phi)$ in powers of $1/L$.
From a straightforward calculation,  one finds up to the order $1/L^2$:
\begin{eqnarray}
\label{eq:UL-expansion}
&& U_L (\phi) \simeq f_L (\beta,\mus) + \frac{\mus}{\beta} \phi  - f_L (\beta,0) \\ 
\nonumber
&& \,\,\, - \frac{1}{\beta L} \ln \sqrt{\frac{2 \pi}{\beta L |f^{(2)}_\star|}} 
+ \frac{1}{\beta L^2} \left[ \frac{f^{(4)}_\star}{8 \beta \mfs^2} + \frac{5 [f^{(3)}_\star]^2}{24 \beta \mfs^3} \right] \, .
\end{eqnarray}
The above equation deserves some comments:

(1) As already mentioned, in the thermodynamic limit, $U_{L \to \infty} (\phi)$ converges to the Gibbs free-energy density $U(\phi)$, which is defined as the Legendre transform of the Helmholtz free-energy density $f_{L \to \infty} ( \beta, h)$ [see Eq.~(\ref{eq:gibbs})] and  is therefore a convex function of the magnetization $\phi$.

(2) Eq.~(\ref{eq:UL-expansion}) is actually an expansion in powers of $\xi/L$. The successive derivatives of the Helmholtz free energy with respect to the external field $\mu$ yield the $n$-points connected correlation functions, $\beta f^{(n)}_\star = (1/L) \sum_{i_1, \ldots , i_n} \langle \varphi_{i_1} \cdots \varphi_{i_1} \rangle\vert_{con}$, which thus behave as $\xi^{n-1}$. As a result,  the expansion of Eq.~(\ref{eq:UL-expansion}) does not converge for $L/\xi < 1$ (even if $L$ is large) and is expected to poorly behave compared to the numerical simulations in this regime. On the other hand, it should provide a good description of the finite-size effective potential for $L/\xi > 1$.

(3) In order to make some use of Eq.~(\ref{eq:UL-expansion}) we need to know the expression of the Helmholtz free energy of the model on a ring of $L$ sites, at temperature $\beta$, and in the presence of an external uniform magnetic field $\mu/\beta$. 

The calculation of  $f_L (\beta,\mu)$ can be done exactly by using a real-space RG approach, called the Migdal-Kadanoff (MK) scheme. It consists in integrating out iteratively half of the sites of the systems (say the odd sites) at each decimation step, and computing recursively the effective pair interaction potential, $W_n (\varphi,\varphi^\prime)$, among the remaining sites. Consider for instance three consecutive sites, $i$, $i+1$, and $i+2$, at the $p$-th step of the renormalization procedure. After integrating out the field on the site $i+1$, one finds the following exact recursive equation:
\begin{eqnarray} \label{eq:MK}
&& W_{p+1} (\varphi_i,\varphi_{i+2}) =  \\
\nonumber 
&& \qquad - \frac{1}{\beta} \ln \int_{- \infty}^{+ \infty} \textrm{d} \varphi_{i+1}
\, e^{- \beta \left[ W_{p} (\varphi_i,\varphi_{i+1}) + W_{p} (\varphi_{i+1},\varphi_{i+2}) \right ] }\, ,
\end{eqnarray}
with the initial condition:
\begin{eqnarray} \label{eq:RG-init}
W_0 (\varphi,\varphi^\prime) &=& \frac{c}{2 a^2} \left( \varphi - \varphi^\prime \right)^2  \\
\nonumber 
&&+ \frac{1}{2} \left[ V(\varphi) + V(\varphi^\prime) 
-\frac{\mu}{\beta} \left( \varphi + \varphi^\prime \right) 
\right] \, .
\end{eqnarray}
For a system of size $L=2^p$, after $p-1$ decimation steps, there are only two sites left and the Helmholtz free-energy density can be obtained as a simple integration:
\begin{equation}
f_L (\beta, \mu) = - \frac{1}{2^p \beta} \ln \int_{- \infty}^{+ \infty} \textrm{d} \varphi \, 
\textrm{d} \varphi^\prime \, e^{- 2 \beta W_{p-1} (\varphi, \varphi^\prime)} \, .
\end{equation}
This procedure allows one to obtain very accurate numerical values of $f_L (\beta, \mu)$ and of its derivatives, provided that the size of the system is an integer power of $2$. In order to access other values of the system size $L$,  we have complemented the RG calculation by a transfer-matrix (TM) approach.

Indeed, the partition function of the system can be written as:
\begin{eqnarray}
\nonumber
Z_L(\beta,\mu) &=& \tr_{\{ \varphi_i \}} T_{\varphi_1,\varphi_2} T_{\varphi_2,\varphi_3} \cdots T_{\varphi_L,\varphi_1} \\
&=& \tr \mathbf{T}^L = \lambda_1^L + \lambda_2^L + \ldots \, ,
\end{eqnarray}
where the transfer-matrix operator is such that $T_{\varphi,\varphi^\prime}=\exp(-\beta W_0 (\varphi,\varphi^\prime))$ with $W_0$ given by Eq.~(\ref{eq:RG-init}). One can then numerically diagonalize the operator by discretizing the values of the fields $\varphi$ and $\varphi^\prime$ and compute its eigenvalues, $\lambda_1, \lambda_2, \ldots$, which leads to an approximate expression for the Helmholtz free-energy density,
\begin{eqnarray} \label{eq:transfer}
\nonumber
f_L (\beta, \mu) &\simeq& - \frac{1}{\beta} \ln \lambda_1 - \frac{1}{\beta L} \, e^{L \ln (\lambda_2 / \lambda_1)} \\
&& \qquad + o \left [ \left( \lambda_3/\lambda_1 \right)^L \right] \, .
\end{eqnarray}
Since the correlation length of the system is given by
\begin{equation}
\xi^{-1} = - \ln \left( \lambda_2 / \lambda_1 \right) \, ,
\end{equation}
Eq.~(\ref{eq:transfer}) provides a good approximation for $f_L (\beta, \mu)$ only for $L \gtrsim \xi$.

\begin{figure}
{\includegraphics[scale=0.65]{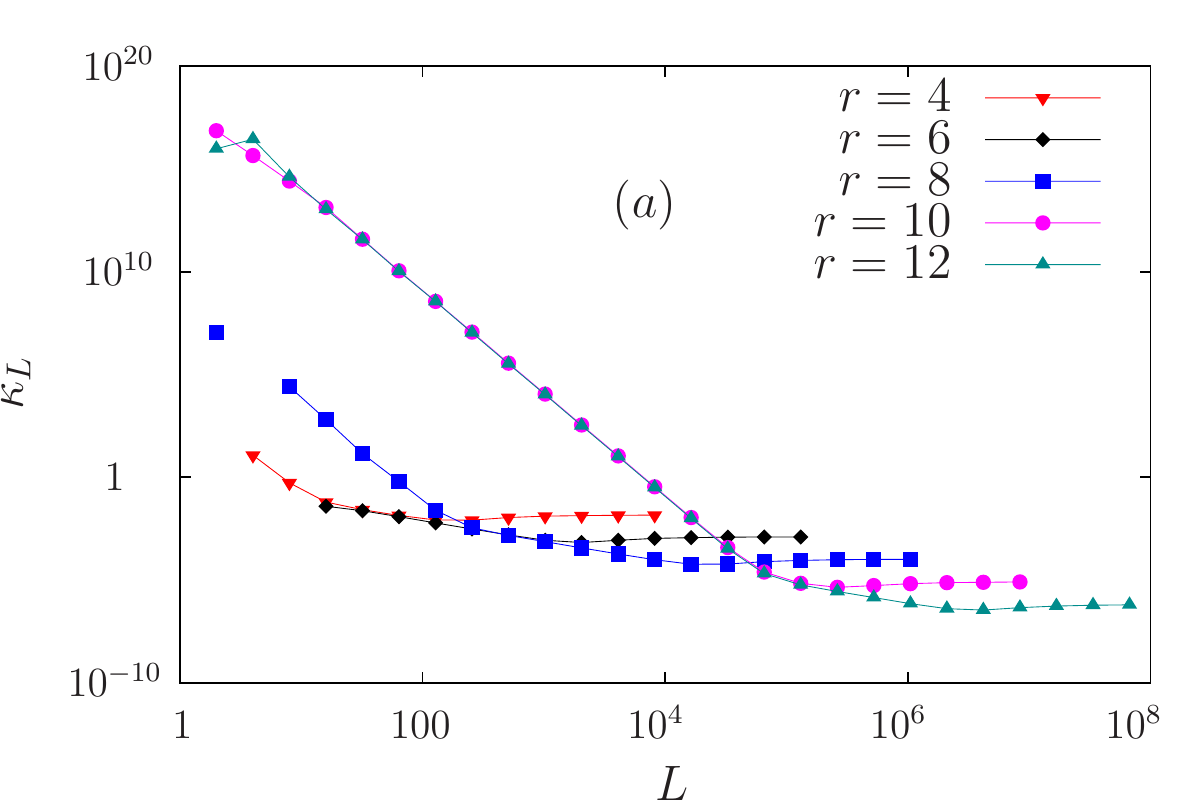}}
{\includegraphics[scale=0.65]{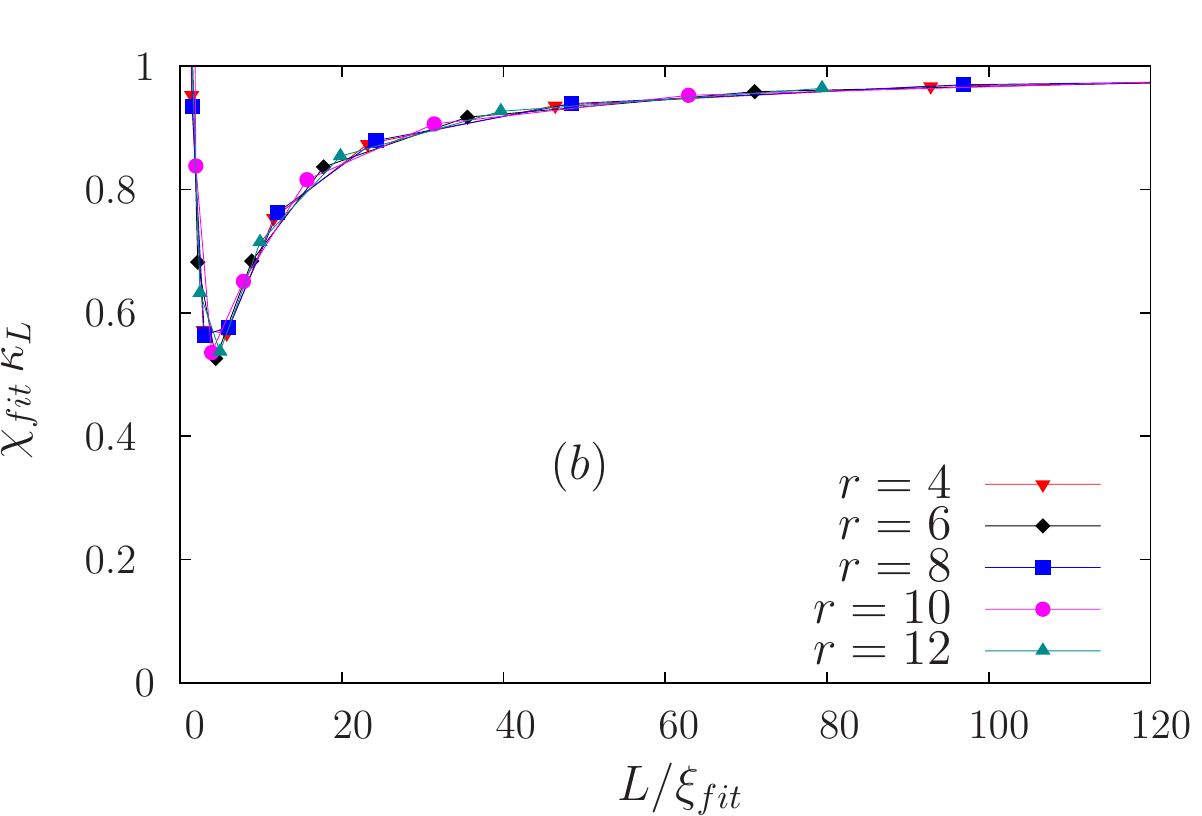}}
\caption{(a): Log-log plot of the curvature $\kappa_L=U''_L(0)$ as a function of $L$ for $\beta=1$, $r= 4$, $6$, $8$, $10$, $12$, and $c=2 r$. (b): Same data with a rescaling of the $x$ and $y$ axes, as $L/\xi_{fit}$ and $\chi_{fit}\kappa_L$ respectively,  to provide the best collapse to a mastercurve.} 
\label{fig:kappa-MC}
\end{figure}

\begin{figure}
{\includegraphics[scale=0.65]{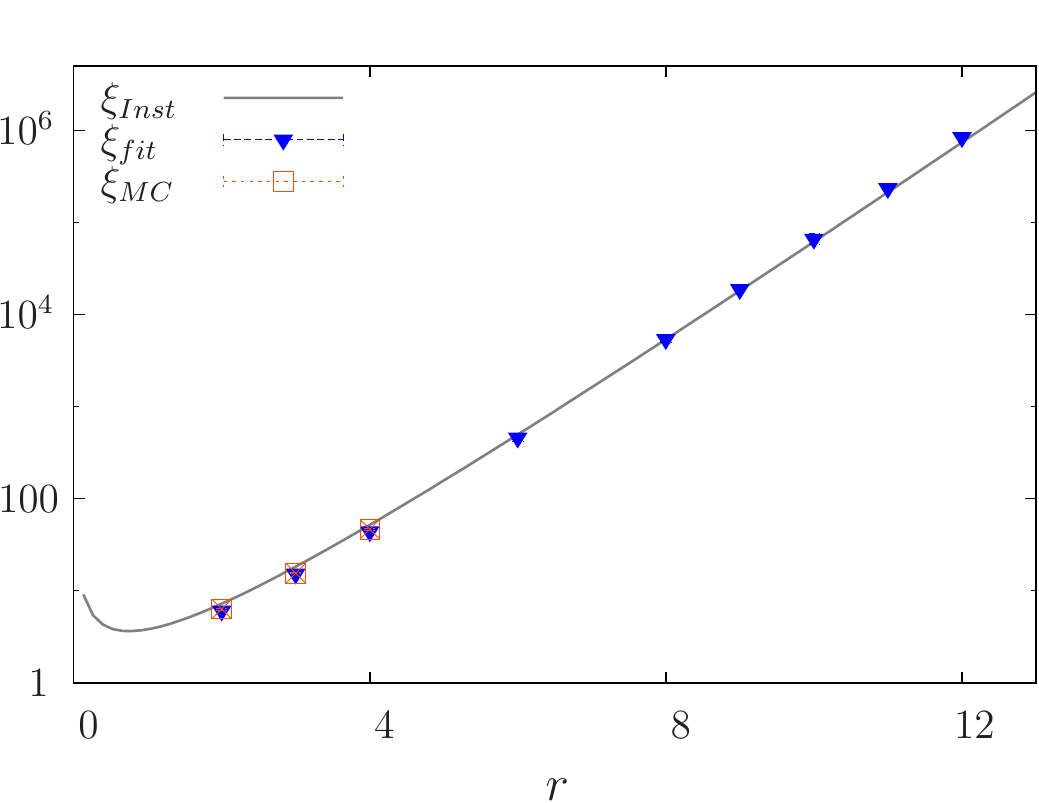}}
\caption{Plot of the scaling parameter $\xi_{fit}$ versus $r$ and comparison with the direct computation of the correlation length through MC and instanton techniques. Recall that have set $\beta=1$, so that the temperature dependence is controlled by $r$. The error bars associated with the fitting procedure are very small in this case and of the order of the symbol size.} 
\label{fig:xi-vs-r}
\end{figure}

\begin{figure}
{\includegraphics[scale=0.65]{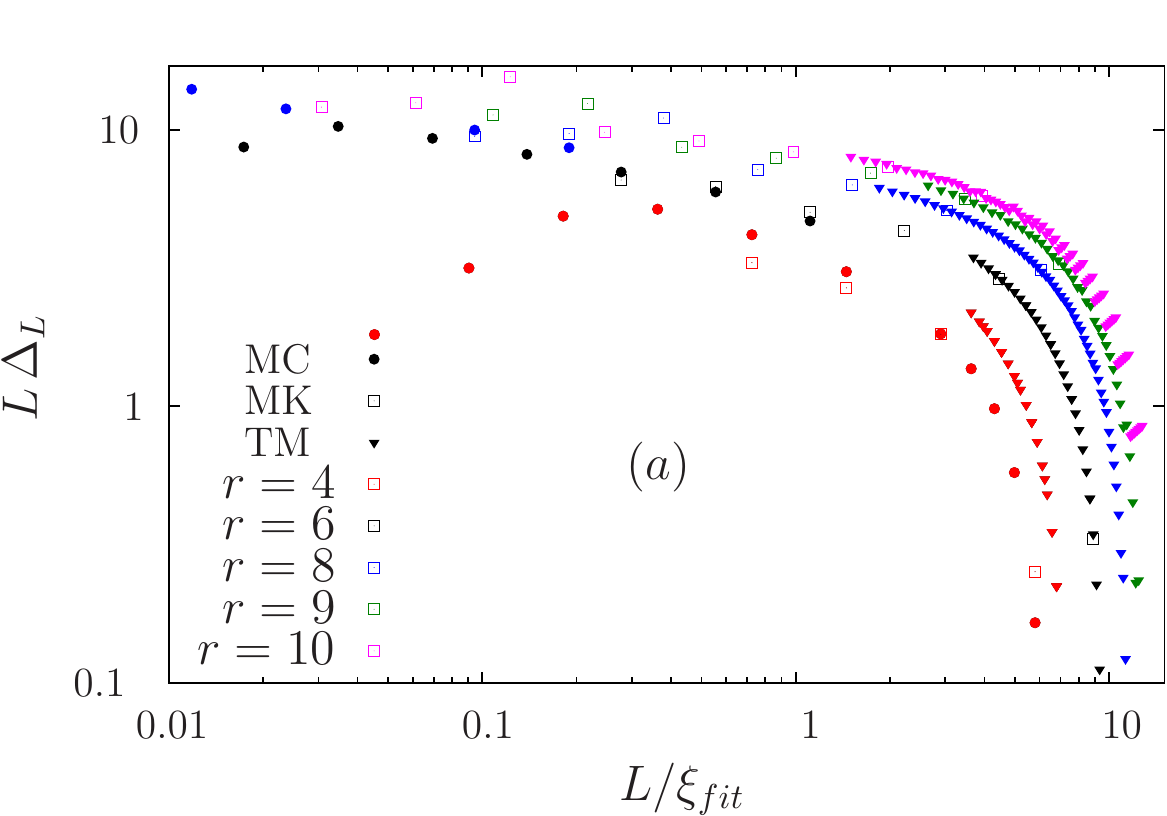}}
{\includegraphics[scale=0.65]{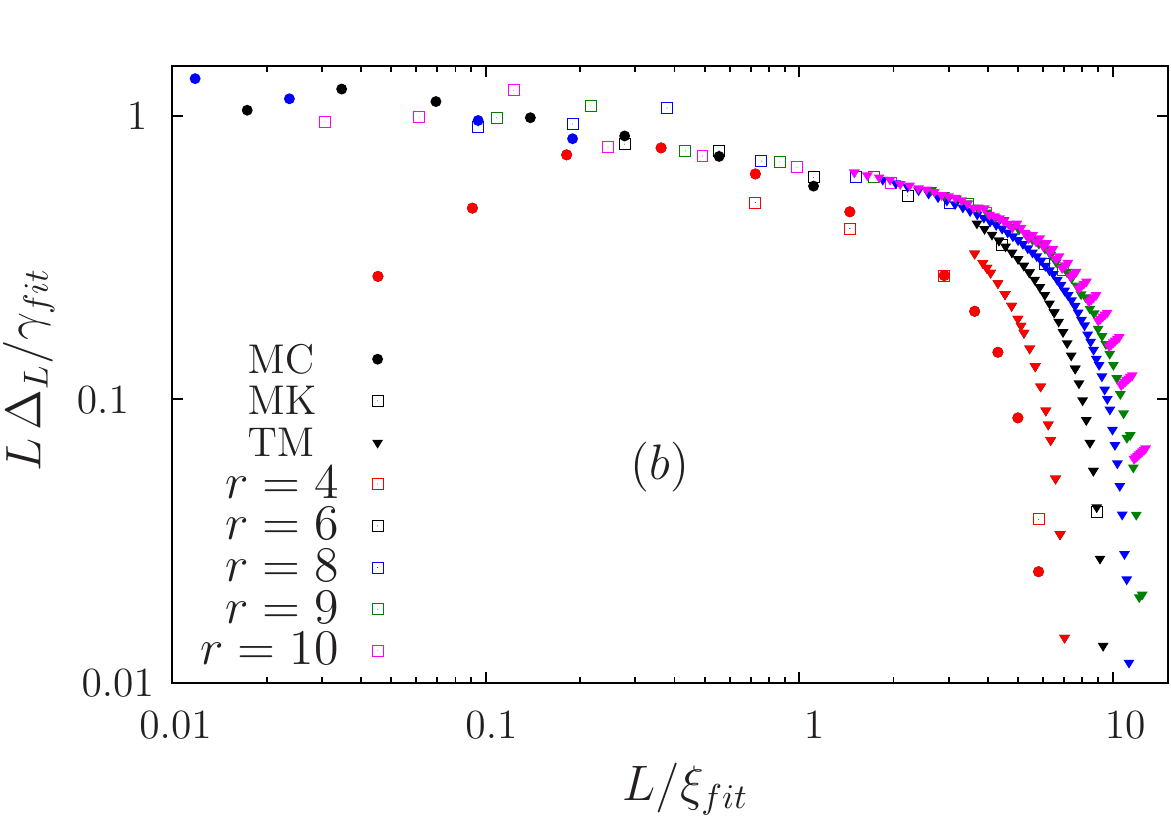}}
\caption{(a): Log-log plot of the barrier height times the system size, $L \Delta_L $, versus $L/\xi_{fit}$ for $\beta=1$,  $c=2 r$, $r= 4$, $6$, $8$, $9$, $10$. Filled circles correspond to MC data. Empty squares are obtained using the $1/L$ expansion with the real-space  RG approach and filled triangles correspond to the $1/L$ expansion with the transfer-matrix approach. 
(b): Same data with an adjustment of the $y$ axis,  $L \Delta_L/\gamma_{fit}$, to provide the best collapse for $L/\xi_{fit}< 1$.} 
\label{fig:delta-MC}
\end{figure}

The finite-size effective potential $U_L(\phi)$ is then obtained from Eq.~(\ref{eq:UL-expansion}). The numerical results for the curvature $\kappa_L$ in $\phi=0$ and for the barrier height $\Delta_L$ at small but finite temperature (or rather, correlation length) are displayed in Figs.~\ref{fig:kappa-MC} and \ref{fig:delta-MC}. In Fig.~\ref{fig:kappa-MC}a, we plot $\kappa_L$ versus $L$ for several temperatures (actually, values of $r$ as we fix $\beta=1$) and in Fig.~\ref{fig:kappa-MC}b we show the best data collapse on a mastercurve after rescaling both the curvature and the system size by temperature-dependent adjustable parameters. (Note that the curvature $\kappa_L$ is obtained from the $1/L$  expansion only as the numerical accuracy of our MC data is not high enough to allow a good determination of  the curvature.) In Fig.~\ref{fig:xi-vs-r} we plot the best-fit parameter $\xi_{fit}$  versus $r$ and compare it to a direct determination of the correlation length through MC simulations and the instanton technique: we find a very good agreement between the two sets of data. The same agreement is obtained for $\chi_{fit}$ which is  found proportional to the correlation length, $\xi_{fit}$ or $\xi$, as expected in one dimension.

\begin{figure}
{\includegraphics[scale=0.65]{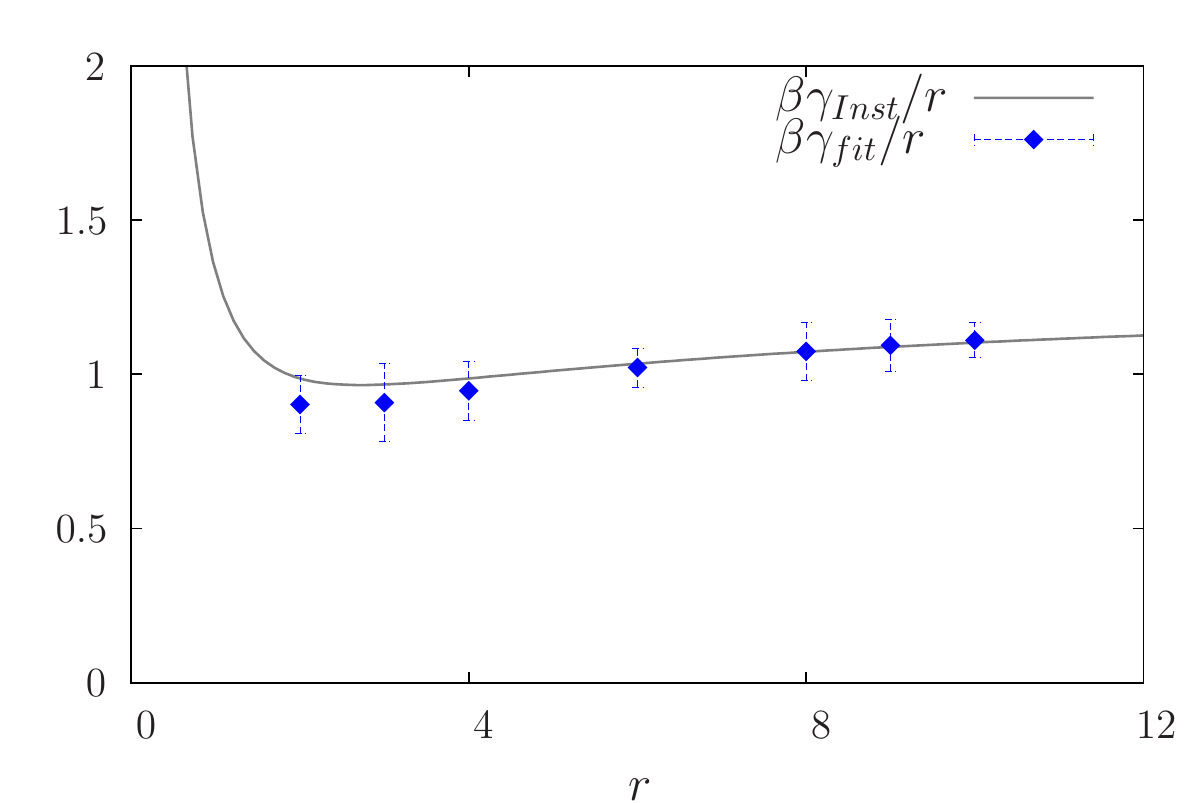}}
\caption{Plot of the scaling parameter $\gamma_{fit}$, multiplied by $\beta/r$, versus $r$ and comparison with the direct instanton computation of the surface tension. The error bars are associated to the uncertainty of the collapse procedure. Recall that we have set $\beta=1$ and $c=2r$, so that the temperature dependence is controlled by $r$, with $\gamma\propto r$. Note also that the instanton calculation breaks down at small $r$.} 
\label{fig:S-vs-r}
\end{figure}

In Fig.~\ref{fig:delta-MC}a, we display a log-log plot of $L \Delta_L$ versus $L/\xi_{fit}$ where $\xi_{fit}$ is obtained from the previous data collapse in Fig.~\ref{fig:kappa-MC}. (As could be anticipated, the $1/L$ expansion fails completely for $L/\xi<1$ and is not shown here.) Fig.~\ref{fig:delta-MC}b then shows the same data with $L \Delta_L$ divided by a temperature-dependent adjustable parameter $\gamma_{fit}$ that ensures the best collapse of all curves for $L/\xi <1$ (this parameter is determined up to a multiplicative constant). When plotted as a function of $r$, we find that this best-fit parameter $\gamma_{fit}$ matches very well the $r$ dependence of the direct estimate of the surface tension of the model through the instanton technique: see Fig.~\ref{fig:S-vs-r}. Here, $\beta=1$ and $c=2r$, so that $\beta  \tilde S^\star= 2r/3-(1/2) \ln r$ for large enough $r$. We have arbitrarily adjusted the unknown constant in $\gamma_{fit}$ so that the latter is roughly equal to $2\beta \tilde S^\star$: the plot in Fig.~\ref{fig:delta-MC}b is shown with this choice of constant (which merely shifts all curves by a constant amount on the log scale).

These plots therefore confirm that the ideas and recipes we have proposed to extract the correlation length $\xi$, the susceptibility $\chi$  and the surface tension $\gamma$ (or alternatively the amplitude of the gradient term $c$) from finite-size numerical data for the effective potential work nicely. Without \textit{a priori} knowledge one can empirically determine the relevant parameters of the underlying (effective) theory from observations on finite-size systems.

\section{Nonperturbative RG}

In this section we take a different approach than the above one. It is more of a ``bottom-up'' approach where we start from a known microscopic (or to the least effective) theory and try to include the fluctuations, in particular possibly strongly nonperturbative ones, to describe the observed macroscopic behavior. As already stated, the method of choice to achieve this is the RG, more precisely the nonperturbative RG (NPRG).
 
The NPRG emerged from Wilson's work\cite{wilson} in the early 70's and has since been formulated is several alternative approaches.\cite{WH,Po,wetterich93}  All of them have the common denominator of treating and summing up fluctuations in a \emph{continuous} way. In this work we focus on the formalism that has been originally developed by Wetterich and coworkers since the 90's.\cite{wetterich93,wetterichreview} In a nutshell, the idea is to start with a given field-theoretical model described by a microscopic (bare) action $S[\varphi]$\cite{footnote_dupuis} and to add to it an infrared (IR) regulator in the form of a mass term:
\beq
S[\varphi]\to S_k[\varphi]=S[\varphi]+\frac 12 \int_q \varphi(q)R_k(q)\varphi(-q)
\eeq
where $\int_q \equiv \int dq/(2\pi)^D$ with $D$ the space dimension and $k$ is a running (momentum) scale. The IR cutoff function $R_k(q)$ goes to zero when $kÊ\to 0$ and provides a mass to the small-momentum modes, $R_k(q) \sim b_k k^2$ when $q\to 0$, but is otherwise arbitrary. One also has that $k\leq \L$ where $\L^{-1}\ll 1$ is the shortest wavelength on which the field $\varphi$ can fluctuate; $\Lambda$ is the ultraviolet (UV) cutoff scale where the continuum theory meets the microscopic details. 

From the regularized action  $S_k[\varphi]$ one can define a cutoff dependent generating functional of the connected correlation function (the analog of a Helmholtz free-energy functional at the scale $k$),
\beq
W_k[J]=\ln\int \mathcal D \varphi \exp\left[-\beta S_k[\varphi]+\int_xJ(x) \varphi(x)\right]\,,
\eeq
where $\int_x \equiv \int d^D x$, and its Legendre transform,
\beq\label{Legendre}
\G_k[\phi]+W_k[J]=\int_x J(x)\phi(x)-\frac \b2 \int_q \phi(q)R_k(q)\phi(-q)\,,
\eeq
where for convenience one subtracts the contribution from the regulator in the definition of $\G_k[\phi]$. The latter is called the effective average action or the running effective action. In the above transformation, $J(x)$ is fixed by the condition that $\phi(x)=\langle \varphi(x) \rangle_k$ and the average is taken by using the modified action $S_k[\varphi]$. 

The running effective action $\G_k[\phi]$ continuously interpolates between the bare action at the UV scale\cite{footnote_UV} and the exact effective action (or Gibbs free-energy functional) $\Gamma[\phi]$, which is the generating functional of the 1PI correlation function, when $k\to 0$. Its evolution with $k$ is described by an exact RG flow equation\cite{wetterich93}
\beq\label{Wett_eq}
\frac{\partial \G_k[\phi]}{\partial k}=\frac \b2 \int_{xy} R_k(x-y)\left[\left(\G_k^{(2)}+\b  R_k\right)^{-1}\right]_{xy}\\
\eeq
with the initial condition $\G_\L[\phi]=\b S[\phi]$ and  $\Gamma_k^{(n)}(x_1,\ldots, x_n)\equiv \delta^{n}\G_k/\delta\phi(x_1)\ldots \delta\phi(x_n)$. By differentiation, this functional flow equation is equivalent to an infinite hierarchy of coupled flow equations for the running effective potential, $U_k(\phi)=\Gamma_k[\phi]/L^D$, and the running 1PI correlation functions (then all evaluated for uniform field configurations).

Finding the exact solution of the functional integro-differential equation in Eq. (\ref{Wett_eq}) is an impossible task in general and one  needs to develop approximation, or closure, schemes that basically replace Eq. (\ref{Wett_eq}) by a finite set of coupled equations for functions. This has been systematically and successfully pursued for a series of problems in both high- and low-energy physics.\cite{wetterichreview}

Applications of the NPRG formalism to the one-dimensional $\varphi^4$ theory with $S[\varphi]$ given by Eq. (\ref{action}) have been previously considered.\cite{TK00,Ao02,Za01,weyrauch06}. In these studies it was found that simple approximation schemes fail to recover the low-temperature physics of the model, in particular the activated scaling of the correlation length. Here, we will show what is the underlying reason for this failure. To this end, we will first derive the exact asymptotic low-temperature form of the running effective action $\Gamma_k[\phi]$ by using the instanton approach and a mapping to the one-dimensional Ising model.

\subsection{Running effective action and instantons in the limit $T\to 0$ ($\xi\rightarrow \infty$)} 
\label{sec:running_instantons}

We first compute the expression of the running effective potential, $U_k(\phi)=\Gamma_k[\phi]/L$ with $\phi$ a uniform field, by using the instanton technique in the low-temperature regime where the correlation length is large (see also above).
     
For the one-dimensional $\varphi^4$ theory under study, the $k$-dependent regularized action reads
\beq
\label{eq:regularized_action}
\begin{split}
S_k[\varphi]&=\int_{0}^{L} \textrm{d} x \left[\frac c 2 (\partial_x \varphi)^2+V(\varphi(x))\right] \\
&+\frac 12 \int_{0}^{L}\de x \int_{0}^{L}\de y\, \phi(x) R_k(x-y)\phi(y)\\
&-L\left[V(\phi_{0,k})+\frac 12 R_k(0)\phi_{0,k}^2\right]
\end{split}
\eeq
where $V(\varphi)$ is given in Eq. (\ref{bare_potential}),
\beq
\phi_{0,k}=\textrm{argmin}_{\varphi}\left[V(\varphi)+\frac 12 R_k(0)\varphi^2\right] \,,
\eeq
and we have added the last term in Eq. (\ref{eq:regularized_action}) for convenience, so that $S_k[\phi_{0,k}]=0$. Contrary to the previous section on the finite-size effective potential, we take here the thermodynamic limit and let $L\to \infty$. The restriction to the spatial extent of the fluctuations is now provided by the IR regulator $R_k(q)$.

A very simple regulator is the Callan-Symanzik one, $R_k(q)=k^2$, which amounts to adding a conventional mass term to the bare action. (Note that in this case the running effective action is equal to the bare action only in the limit $\Lambda \to \infty$ but this has no consequences for the physics at intermediate and small momentum scales.) It is then easy to see that there exists a threshold $k=k_c$ such that $\phi_{0,k}\neq 0$ for all $k\leq k_c$. This threshold corresponds to the moment along the RG flow where the running modified potential $V(\varphi)+\frac 12 k^2\varphi^2$ develops two minima and has a double-well shape. One can expect that this qualitative evolution is very general and does not depend on the details of the regulator. The precise form of $R_k(q)$ changes only the point $k_c$ where the double-well shape first appears.

For $k<k_c$ we can thus evaluate the probability of finding a particular magnetization in the system by using the instanton method, much like in section \ref{sub:finite-size_instantons}.  This probability is given by
\beq
\begin{split}
&P_k(\phi=M/L)=\mathcal N \sum_{n\geq 1}e^{-2n\beta S_k^\star}\int_0^\infty \left(\prod_{i=1}^{2n}\de z_i\right)\\
&\delta\left(\sum_{i=1}^{2n}z_i-L-2n\s_k\right)\delta\left[\sum_{i=1}^{n}\left(z_{2i-1}-h_{2i}\right)-\frac{M}{\phi_{0,k}}\right]
\end{split}
\eeq
where $\mathcal N$ is a normalization constant, $S_k^\star$ is the action evaluated on an single instanton profile and $\s_k$ is the instanton width. By exponentiating  the Dirac delta functions and passing from a discrete sum to a continuum one so that $\a L=2n$, we get
\beq
\begin{split}
&P_k(\phi)=\mathcal N' \int_0^{\infty}\de \a\, \mathrm e^{-\a L \b S_k^{\star}} \int_{-i\infty}^{i\infty} \de \mu\de \nu\int_0^\infty \left(\prod_{i=1}^{2n}\de z_i\right)\\
&\exp\left[\mu L(1+\a \s_k)+L\nu \frac{\phi}{\phi_{0,k}}-\sum_{i=1}^n\left[(\mu+\nu)z_{2i-1}+(\mu-\nu)z_{2i}\right]\right]
\end{split}
\eeq
where $\mathcal N'$ is another normalization constant. Integration over the variables $z_i$ then leads to
\beq
\begin{split}
&P_k(\phi)=\mathcal N' \int_0^\infty \de \a e^{-\a L\b S_k}\int_{-i\infty}^{i\infty} \de \mu\de \nu\\
&\exp\left[L\left(\m+\a\m\s_k+\n\frac{\phi}{\phi_{0,k}} -\frac{\a}{2}\ln(\m^2-\n^2)\right)\right]\:.
\end{split}
\eeq
For large $L$ we can use a saddle point evaluation of the integrals over $\mu$ and $\nu$, which gives
\beq
\begin{split}
&P_k(\phi)=\mathcal N' \int_0^\infty \de \a \exp\Big[-\a L\b S_k\\&
+L\left(\a-\a\ln \a+\frac{\a}{2}\ln[(1+\a\s_k)^2-(\frac{\phi}{\phi_{0,k}})^2]\right)\Big]\:.
\end{split}
\eeq
Finally the integral over $\a$ can also be evaluated with the saddle point method and the value of $\a$ at the saddle point 
is found to be
\beq
\a=\sqrt{1-\frac{\phi^2}{\phi^2_{0,k}}} \mathrm e^{-\beta S_k^\star}
\eeq
where the range of magnetizations is limited to $\phi<\phi_{0,k}$. The low-temperature regime corresponds to $\beta S_k^\star$ large and the system is then described by a dilute instanton gas. The running effective potential $U_k$ is just $-1/(\beta L)$ times $\ln P_k(\phi)$, to which one subtracts the contribution of the IR regulator, and it is given by
\beq\label{U}
\begin{split}
&U_k(\phi)=-\frac{1}{\b L}\ln P_k(\phi)-\frac 12 R_k(0)\left(\phi^2-\phi_{0,k}^2\right)+V(\phi_{0,k})\\
&=-\frac 1\b \sqrt{1-\frac{\phi^2}{\phi_{0,k}^2}}\mathrm e^{-\b S_k^{\star}}-\frac 12 R_k(0)\left(\phi^2-\phi_{0,k}^2\right)+V(\phi_{0,k})\,,
\end{split}
\eeq
which is valid for $\phi<\phi_{0,k}$ and is asymptotically exact when the temperature goes to zero. The associated flow equation reads
\beq\label{localP}
\begin{split}
\frac{\partial U_k(\phi)}{\partial k}&=\frac{\partial}{\partial k}\left[-\frac 1\b \sqrt{1-\frac{\phi^2}{\phi_{0,k}^2}}\mathrm e^{-\b S_k^{\star}}\right]\\
&+\frac 12 \frac{\partial R_k(0)}{\partial k}\left(\phi_{0,k}^2-\phi^2\right)
\end{split}
\eeq

To go further and find the asymptotic low-temperature expressions for the 1PI vertices at scale $k$, we use a short-cut provided by an approximate mapping when $T\to 0$ between the $\varphi^4$ theory and the Ising model. The latter is described by the Hamiltonian
\beq
H[\{\sigma_i\}]=- J\sum_{i=1}^L\s_i\s_{i+1}-h\sum_{i=1}^L\s_i
\eeq
where $\s_{L+1}\equiv\s_1$ if periodic boundary conditions are used and we have set the lattice spacing to one. The thermodynamic limit with $L\to \infty$ is considered here. This calculation for the one-dimensional Ising model is rather standard\cite{baxter07} and the details are given in Appendix~\ref{app:instantons_ising}.

From the comparison between the expressions of the pair correlation function obtained in the low-temperature limit, where the long-distance properties of both theories are described in the continuum,  \textit{i.e.}, for the Ising case
\beq
G_c^{(2)}( r;m)\simeq (1-m^2)e^{-r/\xi(m)}\:.
\eeq
with the correlation length
\beq
\xi(m)\simeq \frac 12 \sqrt{1-m^2} e^{2\beta J}\,,
\eeq
where $m$ is the magnetization per site, and for the (modified) $\varphi^4$ theory (from the instanton calculation),
\beq
G_c^{(2)}( r;\phi)\simeq (\phi_{0,k}^2-\phi^2)e^{-r/\xi(\phi)}
\eeq
where
\beq
\xi(\phi)\simeq \frac 12 \sqrt{1-\frac{\phi^2}{\phi_{0,k}^2}}e^{\beta S_k^\star}\,,
\eeq
one can see that the two theories map onto each other with the following formal replacements when $T\to 0$:
\beq
2J\to S_k^\star \ \ \ \ \ \ \ \ \ \ (1-m^2)\to (\phi_{0,k}^2-\phi^2)
\eeq
and $S_k^\star$, the instanton action, is identified with the domain-wall energy when $T\to 0$.

At low temperature, the two-point 1PI correlation function can be written as
\beq
\frac 1\b\G^{(2)}_k(p;\phi)\simeq \frac{1}{2\b(\phi_{0,k}^2-\phi^2)}\xi(\phi)\left[p^2+\xi^{-2}(\phi)\right]-R_k(0)
\eeq
where the last term is due to the definition of the running effective action in Eq. (\ref{Legendre}). This expression can be put in the form\beq\label{Gamma2}
\beta^{-1} \Gamma_k^{(2)}(p;\phi)= Z_k(\phi)p^2+U''_k(\phi)
\eeq
where $Z_k(\phi)$ can be obtained as
\beq
\label{eq:Zdef}
Z_k(\phi)=\lim_{p\to 0}\frac{1}{2\b}\frac{\partial^2}{\partial p^2}\G_k^{(2)}(p,\phi)=\frac{1}{2\b(\phi_{0,k}^2-\phi^2)}\xi(\phi)\,.
\eeq
In addition, we can check that
\beq
\label{eq:Uprimeprime}
U''_k(\phi)+ R_k(0)=\frac 1\b\frac{1}{\phi_{0,k}^2-\phi^2}\frac{e^{-\b S_k}}{\sqrt{1-\phi^2/\phi_{0,k}^2}}\,,
\eeq
in complete agreement with Eq. (\ref{U}).

By using the mapping with the one-dimensional Ising model, we can also obtain low-temperature expressions for the higher-order 1PI correlation functions, $\Gamma^{(3)}_k$, $\Gamma^{(4)}_k$, $\cdots$, at the scale $k$. With the results given in Appendix~\ref{app:1PIvertices_ising} we obtain
\beq
\begin{split}\label{Gamma3}
\Gamma^{(3)}_k &(p_1,p_2,p_3;\phi)=(2\pi)\delta(p_1+p_2+p_3)\frac{c(\phi)s(\phi)^2}{2\xi(\phi)}\\
&\times\left[3-\xi(\phi)^2\left(p_1p_2+p_1p_3+p_2p_3\right)\right]
\end{split}
\eeq

\beq\label{Gamma4}
\begin{split}
&\G_k^{(4)}(p_1,p_2,p_3,p_4;\phi)=(2\pi)\d(p_1+p_2+p_3+p_4) \frac{1}{2\xi(\phi)s(\phi)^6}\\
&\times \left(-[c(\phi)^2+s(\phi)^2]\xi(\phi)^4p_1p_2p_3p_4 -[3c(\phi)^2+s(\phi)^2] \times \right.\\
&\left.\xi(\phi)^2 (p_1p_2+p_1p_3+p_1p_4+p_2p_3+p_2p_4+p_3p_4)\right.\\ &\left.+3[5c(\phi)^2+s(\phi)^2]\right)
\end{split}
\eeq
where $c(\phi)=\phi$ and $c(\phi)^2+s(\phi)^2=\phi_{0,k}^2$ ($c$ should not be confused with the notation also used for the prefactor of the derivative term in the bare action). More generally the running 1PI correlation function can be cast in the form
$\G_k^{(n)}(p_1,\cdots,p_n;\phi)=(2\pi)\delta(p_1+\cdots+p_n) \xi(\phi)^{-1}g_n(\xi(\phi)p_1,\cdots,\xi(\phi)p_n;\phi)$ where the remaining $\phi$ dependence in $g_n$ does not contain exponential terms involving $\exp(\beta S_k^\star)$.

\subsection{Approximation schemes}

Having obtained the exact expressions for the running effective potential and the running 1PI correlation functions in the low-temperature limit, we can now test approximation schemes for the exact NPRG equation in Eq. (\ref{Wett_eq}). Among the several approximation schemes so far proposed, we will focus first on the most popular one, the so-called derivative expansion. In this approximation, the running effective action at the scale $k$ is expanded in gradients of the field,
\beq
\Gamma_k[\phi]= \beta\int_x \left[U_k(\phi(x)) + \frac 12 Z_k(\phi(x))\left(\frac{\partial \phi(x)}{\partial x}\right)^2+ \cdots \right]
\eeq
where the higher-order terms involve $4$, $6$, etc., derivatives of the field. We will show that finite truncations of the derivative expansion are unable to reproduce the exact features of the low-temperature physics.

\subsubsection{LPA} 

The Local potential Approximation (LPA) is the lowest order of the derivative equation. It corresponds to
\beq
\Gamma_k[\phi]= \beta\int_x \left[U_k(\phi(x))+\frac 12\left(\frac{\partial \phi(x)}{\partial x}\right)^2\right]\,,
\eeq
where the coefficient of the gradient term is constant and is not renormalized. Plugging this ansatz into Eq. (\ref{Wett_eq}), computing it for a uniform field $\phi$ and choosing the simple regulator $R_k(q)=b_k k^2$ with $b_k$ constant (taken to $1$) lead to the following differential equation for the running effective potential $U_k(\phi)$:
\beq\label{LPA_eq}
\partial_kU_k(\phi)=\frac{1}{4\b}\partial_k( k^2)\frac{1}{\sqrt{U''_k(\phi)+ k^2}}\,.
\eeq
It is easily verified that the exact expression in Eq. (\ref{U}) does not satisfy the above equation. The latter is actually unable to reproduce the correct scaling of the correlation length, with, \textit{e.g.}, $U_k''+k^2\propto \exp(-\beta S_k^{\star})$ [see Eq. (\ref{eq:Uprimeprime})].

In Fig. \ref{fig:Gamma(m)}, we have plotted the running effective potential $U_k(\phi)$ at several values of $k$, as obtained from the LPA with a regulator of the form\cite{litim01} $R_k(p)=(k^2-p^2)\Theta(k^2-p^2)$. The curves illustrate the return to convexity of the potential. However, as also known from previous attempts,\cite{TK00,Ao02,weyrauch06} if the LPA provides a good description for values of $T$ higher than the energy barrier of the double well, or more precisely than the instanton energy cost $S^\star$, they fail to reproduce the low-temperature result with a thermally activated dependence of the correlation length, $\propto \exp(\beta S^\star)$. For instance, the curvature of the effective potential in zero, $\kappa_{k=0}=U''_{k=0}(0)$, which should vanish exponentially when $T \to 0$ as $\exp(-\beta S^\star)$ (see also section \ref{sec:U_L}) is generically found to vanish as a power law of $T$ instead. The nonperturbative regime associated with the rare localized events, which is captured by the instanton calculation, is therefore completely missed.

\subsubsection{Second order of the derivation expansion}

The next order corresponds to the following ansatz
\beq
\G_k[\phi]=\b\int_x\left[U_k(\phi(x))+ \frac 12 Z_k(\phi) \left(\frac{\partial \phi}{\partial x}\right)^2\right]\:.
\label{ansatz}
\eeq
\begin{widetext}
When inserted in the exact RG flow equation, this ansatz leads to two coupled differential equations for the functions $U_k(\phi)$ and $Z_k(\phi)$ [the latter is obtained from the exact flow equation for the second vertex $\Gamma_k^{(2)}$ with the use of the prescription given in the first equality of Eq. (\ref{eq:Zdef})]:
\beq\label{DE11}
\partial_k U_k(\phi)=\frac{1}{4\b}\partial_k (b_k k^2)\left[Z_k(\phi)\left(U''_k(\phi)+b_k k^2\right)\right]^{-1/2}
\eeq
\beq\label{DE12}
\begin{split}
\partial_kZ_k(\phi)&=\frac 1\beta\partial_k(b_k k^2)\left[-\frac{5}{64}U'''_k(\phi)^2Z_k(\phi)^{1/2}\left(U''_k(\phi)+b_k k^2\right)^{-7/2}+\frac{9}{32}Z_k'(\phi)U_k'''(\phi)Z_k(\phi)^{-1/2}\left(U''_k(\phi)+b_k k^2\right)^{-5/2}\right.\\
&\left.+\frac{7}{64}Z_k'(\phi)^2Z_k(\phi)^{-3/2}\left(U''_k(\phi)+b_k k^2\right)^{-3/2}-\frac 18Z_k''(\phi)Z_k(\phi)^{-1/2}\left(U''_k(\phi)+b_k k^2\right)^{-3/2}\right]
\end{split}
\eeq
where the IR cutoff function is of the same form as above (and a residual $k$-dependence is allowed in $b_k$). 

When inserting the exact expression for $U_k(\phi)$ and $Z_k(\phi)$ given in Eqs. (\ref{U}), (\ref{eq:Zdef}), and (\ref{eq:Uprimeprime}), one can see that Eq. (\ref{DE11}) is now satisfied at leading order in $\exp(-\beta S_k^{\star})$ but not Eq. (\ref{DE12}). The exact expressions indeed generate terms of order $\exp(2\beta S_k^{\star})$ in the right-hand side of Eq. (\ref{DE12}) which do not cancel and have no counterparts in the left-hand side [which is itself essentially of order $\exp(\beta S_k^{\star})$]. The problem found at the LPA level can be formally cured at the level of the effective average potential but at the expense of an inconsistency at the level of the function $Z_k(\phi)$.

\subsubsection{Fourth order of the derivative expansion}

To check whether the results found above correspond to a more systematic pattern, we have considered the fourth order, which corresponds to taking
\beq
\G_k[\phi]=\int_x \left[U_k(\phi(x))+\frac 12Z_k(\phi(x))(\partial\phi(x))^2+ \frac{1}{4!}Y_k(\phi(x))\left(\partial \phi(x)\right)^4\right] \,.
\eeq
The equation for the running effective potential in Eq. (\ref{DE11}) is unchanged but that for $Z_k(\phi)$ is now obtained as
\beq
\begin{split}
\partial_kZ_k(\phi)&=\frac 1\beta\partial_k(b_k k^2)\left[-\frac{5}{64}U'''_k(\phi)^2Z_k(\phi)^{1/2}\left(U''_k(\phi)+b_k k^2\right)^{-7/2}+\frac{9}{32}Z_k'(\phi)U_k'''(\phi)Z_k(\phi)^{-1/2}\left(U''_k(\phi)+b_k k^2\right)^{-5/2}\right.\\
&\left.+\frac{7}{64}Z_k'(\phi)^2Z_k(\phi)^{-3/2}\left(U''_k(\phi)+b_k k^2\right)^{-3/2}-\frac 18Z_k''(\phi)Z_k(\phi)^{-1/2}\left(U''_k(\phi)+b_k k^2\right)^{-3/2}-\frac 18 Y_k(\phi)Z_k(\phi)^{-3/2} \times \right.\\& \left. \left(U''_k(\phi)+b_k k^2\right)^{-1/2}\right] \,.
\end{split}
\label{ZY}
\eeq
\end{widetext}
An equation for $Y_k(\phi)$ is also derived by considering the flow of the $4$-point 1PI vertex but it is too long to be given here. 

When inserting the exact low-temperature expressions for $U_k(\phi)$, $Z_k(\phi)$, and $Y_k(\phi)$ [the latter can be obtained from Eqs. (\ref{Gamma3},\ref{Gamma4})] in the three flow equations corresponding to the present ansatz, one finds that both the equation for $U_k$ and that for $Z_k$ in Eq. (\ref{ZY}) are satisfied. For the latter, the term involving $Y_k(\phi)$ in the right-hand side of Eq. (\ref{ZY}) now exactly cancels the term in $\exp(2\beta S_k^{\star})$ that led to an inconsistency in the second-order approximation (see above). On the other hand, one can check that the approximate equation for $Y_k(\phi)$ is not satisfied by the exact expression because of the presence of terms of order $\exp(4\beta S_k^{\star})$ in the right-hand side [while $Y_k$ itself behaves as $\exp(3\beta S_k^{\star})$].

\subsubsection{General scheme and further approximations}

Guided by the above results, it is now easy to infer the general pattern. The prefactors of the terms with $2l$ derivatives of the field in the derivative expansion of the running effective action $\Gamma_k[\phi]$ are of order $\exp[(2l-1)\beta S_k^{\star}]$ in the low-temperature regime [and $U''_k(\phi)+R_k(0)$ is itself of order $\exp(-\beta S_k^{\star})$]. This dominant behavior when $\beta S_k^{\star}\to \infty$ emerges from the exact NPRG hierarchy of equations for the 1PI vertices because terms that would naively lead to a higher power in $\exp(\beta S_k^{\star})$ in the right-hand side of the equations (the ``beta functions'') exactly cancel out. This cancelation effect is however lost if one truncates the expansion, whatever the order of the truncation. We conjecture that the appropriate ansatz of $\Gamma_k[\phi]$ that reproduces the low-temperature physics of the model is instead
\beq\label{general_ansatz}
\G_k[\phi]=\int_x\left[ U_k(\phi(x)) +\sum_{l=1}^\infty\frac{1}{(2l)!}Y_{k,2l}(\phi(x))(\partial \phi(x))^{2l}\right]
\eeq
with, to make contact with the previous notations, $Y_{k,2}(\phi)\equiv Z_k(\phi)$ and $Y_{k,4}(\phi)\equiv Y_k(\phi)$.
Note that the above form of $\Gamma_k$ is \textit{not} the most general one: in the derivative expansion, the term of order $\partial^{2l}$ is actually a combination of terms involving $(\partial \phi)^{2l}$, $\partial^{2}\phi (\partial \phi)^{2l-2}$, $\cdots$, $\partial^{2l}\phi$ which even after integration by part cannot in general be reduced to a single contribution as in Eq. (\ref{general_ansatz}). The specific form in Eq. (\ref{general_ansatz}) results from the rather simple momentum dependence of the 1PI correlation functions in the one-dimensional Ising model and $\varphi^4$ theory at low temperature. 

The above finding allows us to discuss another approximation of the NPRG called BMW.\cite{BMWref} It corresponds to a closure of the exact NPRG hierarchy at the level of the equation for the 1PI two-point function $\Gamma_k^{(2)}(p,\phi)$:
\beq
\begin{aligned}
\label{G2flow}
&\partial_k\G^{(2)}_k(p,\phi)=\b\int\frac{\de q}{2\pi}\partial_kR_k(q)\big[G_k(q,\phi)^2G_k(p+q,\phi) \times 
\\& \G^{(3)}_k(p,q,-p-q;\phi)^2-\frac 12 G_k(q,\phi)^2\G^{(4)}_k(p,-p,q,-q;\phi)\big]
\end{aligned}
\eeq
where $G_k(p,\phi)=[\G^{(2)}_k(p,\phi)+R_k( p)]^{-1}$. The BMW closure consists in setting to zero the internal momentum $q$ appearing in the $3$- and $4$- point vertices in the right-hand side. After using the consistency relations, $ \G^{(3)}_k(p,0,-p;\phi)=\partial\G^{(2)}_k(p;\phi)/\partial \phi$ and $\G^{(4)}_k(p,-p,0,0;\phi)=\partial^2\G^{(2)}_k(p;\phi)/\partial \phi^2$, one obtains a closed equation
\beq
\begin{aligned}
\label{G2flowBMW}
&\partial_k\G^{(2)}_k(p,\phi)=\b\int\frac{\de q}{2\pi}\partial_kR_k(q)\big[G_k(q,\phi)^2G_k(p+q,\phi) \times 
\\& \left [\frac{\partial \G^{(2)}_k(p;\phi)}{\partial \phi}\right]^2-\frac 12 G_k(q,\phi)^2\frac{\partial^2 \G^{(2)}_k(p;\phi)}{\partial \phi^2}\big]
\end{aligned}
\eeq
which can be combined with the equation for the running effective potential $U_k(\phi)$. It is easily checked that Eq. (\ref{G2flowBMW}) is not compatible with the exact low-temperature expressions of $U_k$ and $ \G^{(2)}_k$ given in section \ref{sec:running_instantons}: after scaling the momenta by $\xi(\phi)$ (see section \ref{sec:running_instantons}), the left-hand side of Eq. (\ref{G2flowBMW})  scales as $\xi^{-1}$ whereas the right-hand side has a term in $\xi^0$ that does \textit{not} cancel out. Just like truncations of the derivative expansion, the BMW closure is therefore unable to properly describe the nonperturbative physics of the one-dimensional $\varphi^4$ at low temperature.

The alternative to the existing approximation schemes of the NPRG is to start from the exact low-temperature ansatz in Eq. (\ref{general_ansatz}). This however leads to an infinite set of differential equations that cannot be treated with standard methods. We have tried another route which amounts to considering the running effective action as being local in the two variables $\phi(x)$ and $\partial\phi(x)$ and introduce an auxiliary field $\hat\phi(x)$ to decouple $\partial\phi(x)$ from $\phi(x)$. This procedure however is highly ambiguous. In addition, say we end up with a running effective action of the form $\Gamma_k[\phi,\hat\phi]=\int_x \mathcal V_k(\phi(x),\hat\phi(x))$, it is not clear that standard approximations on this ansatz will correctly capture the expected low-temperature physics. Actually we have tried an LPA approximation at the level of the two fields $\phi$ and $\hat\phi$ and it completely misses the nonperturbative regime. More work is needed to possibly find a solution to this unsatisfactory theoretical situation.

 \section{Discussion and concluding remarks}
 
In this work we have studied the $\varphi^4$ theory at low temperature in the regime where the behavior of the system is completely dominated by nonperturbative instantonic fluctuations. We have first discussed empirical recipes to extract the parameters of the underlying microscopic or effective theory from the numerical study of finite-size systems. This strategy could be very useful in the analysis of finite-size numerical simulation of glassy systems but the application to this problem deserves further work.

Our study also illustrates the difficulty to describe the low-temperature nonperturbative physics of the one-dimensional $\varphi^4$ theory through truncations of the NPRG. In a sense, however, the one-dimensional case is harder than the situation in higher dimensions. There, the transition associated with a spontaneous symmetry breaking is not destroyed by the fluctuations and the return to convexity has been shown to be properly described through simple approximations of the NPRG.\cite{parola-reatto,tetradis92,wetterichreview} We now discuss in more detail this higher-dimensional situation.

\begin{figure}
{\includegraphics[scale=0.3]{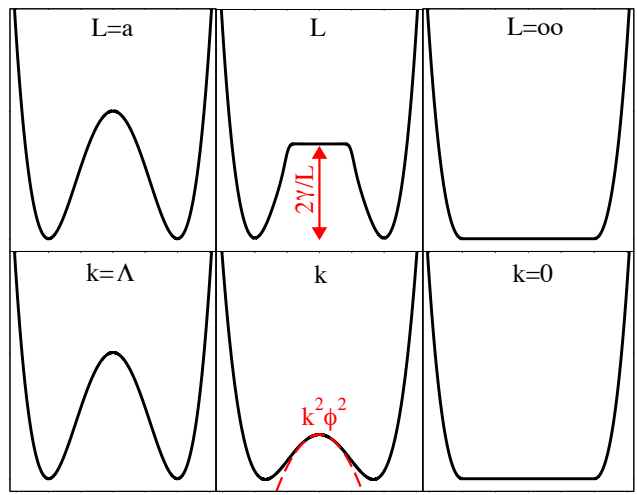}}
\caption{Schematic plot of the evolution of the shape of the finite-size effective potential $U_L(\phi)$ (top) and of the running effective potential $U_k(\phi)$ (bottom) with either system size $L$ or running IR momentum scale $k$ for the $\varphi^4$ theory in $3$ dimensions.} 
\label{fig:schematic_3D}
\end{figure}

Consider for instance the $3$-dimensional $\varphi^4$ theory. As far as the finite-size effective potential $U_L(\phi)$ is concerned, one can repeat and adapt the qualitative arguments developed in section \ref{sec:shape_U_L}. At low enough temperature, the bare potential has two minima in, say, $\phi=\pm 1$ and the relevant excitations above the uniform ground states are system-spanning domain walls or interfaces between regions of essentially constant positive and negative magnetization. When the system size $L$ becomes larger than the interface width, the system can accommodate one system-spanning interface: $U_L(\phi)$ should then have, on top of the two symmetric minima for $\phi \simeq \pm 1$, a plateau for intermediate values of the field; the height of the plateau compared to the bottom of the minima is given by $\Upsilon/L$ where $\Upsilon$ is the surface tension. As $L$ increases, this height decreases and goes to zero in the thermodynamic limit. The effective potential is convex with a flat intermediate portion corresponding to phase coexistence. The evolution with $L$ of $U_L(\phi)$ is schematically depicted in Fig. \ref{fig:schematic_3D}. In this case, studying finite-size systems should  allow one to extract two physical quantities, the surface tension and the correlation length which corresponds to the interface width.

This $3$-dimensional $\varphi^4$ theory in the symmetry-broken region\cite{footnote_spin_waves} has also been studied in detail within the NPRG framework.\cite{parola-reatto,tetradis92,wetterichreview} Although influenced by domain walls as in one dimension, the long-distance physics is nonetheless different as these nonperturbative fluctuations are not strong enough to destroy the phase transition. As a result, simple approximation schemes of the running effective action properly capture the effect of these fluctuations. The running effective potential $U_k(\phi)$ evolves with decreasing $k$ from the bare double-well potential to a convex effective potential when $k=0$: see the schematic plot in Fig. \ref{fig:schematic_3D}. Provided one chooses an appropriate class of IR regulator,\cite{wetterichreview} the intermediate ``inner'' part of $U_k(\phi)$ displays at small $k$ a parabolic shape $\propto k^2\phi^2$ that comes in addition to the two symmetric minima in $\phi=\pm \,\phi_{0,k}$. This parabolic dependence corresponds to the expected exact behavior obtained by considering the nonuniform configurations of the field involving domain walls. The remarkable feature is that this behavior is recovered by using approximations of the NPRG, such as the first orders of the derivative expansion, which only consider expansions about uniform fields.\cite{tetradis92} This is in stark contrast with the situation encountered in one dimension.

The nature of the NPRG flow somehow changes when the running IR momentum scale $k$ crosses some value $k_\star$ that roughly corresponds to the point at which $k^2$ becomes of the order of magnitude of the curvature of the running effective potential in $\phi=0$: $k_\star^2\sim \vert U''_{k_\star}(0)\vert$; this in turn corresponds to the point where $1/k$ becomes of the order of the width of the domain wall in the (nonuniform) field configurations that minimize the running effective action at this scale $k$.\cite{tetradis92,wetterichreview} Whereas such information can be included in an improved instantonic theory of nucleation in the cases where metastability in present,\cite{strumia99,wetterichreview} as, \textit{e.g.}, when applying a nonzero external source or magnetic field, its interpretation in terms of physical quantities of the actual, macroscopic system remains unclear. In particular this length scale $1/k^\star$ does not provide any direct information on one of the important length scales in nucleation problems, \textit{i.e.}, the size of the critical droplet or bubble.

In any case, we think that providing a generic solution to the problem posed by nonperturbative fluctuations in model systems as the one studied here would be very profitable for tackling the harder situations encountered in glassy systems which involve activated dynamics in a complex landscape.

\begin{acknowledgments}
We thank J.-P. Bouchaud, C. Cammarota, L. Canet, B. Delamotte and G. Parisi for fruitful discussions and we acknowledge support from the ERC grant NPRGGLASS.
\end{acknowledgments}

\appendix

\section{Computation of the combinatorial factors for the gas of instantons}
\label{app:instantons}

The combinatorial coefficients $I_{2n}(L)$ are configuration integrals of $2n$ domain walls of width $\sigma$ on a ring of size $L$. This problem is equivalent to the computation of the partition function of $2n$ (discernible) hard spheres of size $\sigma$ on a ring of size $L$ in $D=1$. As done in the main text, we define $x_i$, $i=1, \ldots, 2n$, as the lengths of the regions with constant $\phi = \pm 1$ (\textit{i.e.}, the gaps between the spheres). These variables must satisfy the constraint $\sum_{i=1}^{2n} x_{i}+ 2n\sigma = L$.
\begin{widetext}
One then has 
\begin{displaymath}
I_{2n}(L) = \frac{L}{n} \int_{0}^{L-2n\sigma} \de x_{2n-1} \int_{0}^{L-2n\sigma-x_{2n-1}} \de x_{2n-2} \cdots \int_{0}^{L-2n\sigma - (x_{2n-1}+x_{2n-2} + \cdots + x_2)} \de x_1 \, ,
\end{displaymath}
where the factor $L$ comes from translational invariance and the factor $1/n$ accounts for the number of ways one can choose the
first kink/anti-kink pair. In the following, we will determine the expression of $I_{2n}$ by recurrence. 
In order to do this, it is convenient to introduce the functions
\begin{equation} \label{eq:g}
g_n(y)= \int_{0}^{y} \de x_{n-1} \int_{0}^{y-x_{n-1}} \de x_{n-2} \cdots \int_{0}^{y - (x_{n-1}+ x_{n-2} + \cdots + x_2)} \de x_1 \, ,
\end{equation}
\end{widetext}
in terms of which the combinatorial factors can be expressed as
\begin{equation} 
\label{eq:Ig}
I_{2n}(L) = \frac{L}{n} \, g_{2n} (L - 2 n \sigma) \, .
\end{equation}
From the definition in Eq.~(\ref{eq:g}), one can write $g_{n+1} (y)$ in terms of $g_n (y)$,
\begin{equation} \label{eq:gk1}
g_{n+1} (x) = \int_0^y \de x_n \, g_n (y - x_n) = \int_0^y \de x_n \, g_n (x_n) \, ,
\end{equation}
which, from $g_2(y) = \int_0^y \de x_1 = y$ and by recurrence, immediately leads to
\begin{equation} \label{eq:guess}
g_n (y) = \frac{y^{n-1}}{(n-1)!} \, .
\end{equation}
Finally, after plugging  Eq.~(\ref{eq:guess}) into (\ref{eq:Ig}), one obtains Eq.~(\ref{eq_I_2n}) of the main text.

In order to compute the combinatorial factors $J_{2n} (M,L)$ one has to impose that the gaps $x_i$ satisfy the two following constraints:
\begin{displaymath}
\left \{
\begin{split}
&\sum_{i=1}^{2n} x_{i} = L - 2n\sigma \, , \\
&\sum_{i=1}^n \left( x_{2i-1} - x_{2 i} \right) = M \, ,
\end{split}
\right .
\end{displaymath}
which can be rewritten as
\begin{displaymath}
\left \{
\begin{split}
& \sum_{i=1}^n x_{2i-1} = \dfrac{L - 2k\sigma + M}{2} \, ,\\ 
& \sum_{i=1}^n x_{2i} = \dfrac{L - 2k\sigma - M}{2} \, .
\end{split}
\right .
\end{displaymath}
As a result, the integrals over the variables $x_i$ can be divided into separate integrations over even and odd gaps, 
which can be written in terms of the functions $g_n (y)$ defined above. This yields
\begin{displaymath}
\begin{split}
J_{2n} (M,L) &= \frac{L}{2 n} \, g_n \! \left( \frac{L - 2 n \sigma - M}{2} \right ) \\
& \qquad \qquad \times 
g_n \! \left( \frac{L - 2 n \sigma + M}{2} \right ) \, .
\end{split}
\end{displaymath}
The extra $1/2$ factor comes from the fact that only half of the configurations, namely, those with the first domain wall 
joining $\varphi=-1$ to $\varphi = +1$, contribute to magnetization $+M$, whereas the others contribute to $-M$.
After using the exact expression in Eq.~(\ref{eq:g}), one finally finds
\begin{equation}
J_{2n}(M,L) = \frac{L}{2n} \,
\frac{\left[ \left (L - 2n\sigma \right)^2 - M^2\right]^{n-1}}{2^{2(n-1)} (n-1)!^2} \, ,
\end{equation}
which, with the help of the intensive variables $\phi = M/L$ and $\alpha= \sigma/L$,
leads to Eq.~(\ref{eq:PLm-instantons}) of the main text.

\begin{widetext}
In the following we show that $P_L( \phi)$ given in Eq.~(\ref{eq:PLm-instantons}) is properly normalized to $1$. We start by computing the integrals over $\phi$ of the terms of the sum separately. By changing variable to $x = \phi/(1 - 2 n \alpha)$ one gets
\begin{displaymath}
\Upsilon_{2n}(L) = 2 \left(\frac{\zeta}{2} \right)^{2n} \int_{-(1-2n\alpha)}^{+(1-2n\alpha)} \de \phi \, \frac{\left[ \left(1 - 2n\alpha\right)^2 - \phi^2\right]^{n-1}}{n!(n-1)!}
= 2\,\frac{(\zeta/2)^{2n}}{n!(n-1)!} \, (1 - 2n\alpha)^{2(n-1)} (1-2n\alpha) \int_{-1}^{+1} \de x \, (1 - x^2)^{n-1} \, .
\end{displaymath}
The integral over $\de x$ can be computed as
\begin{displaymath}
\int_{-1}^{+1} \de x \, (1 - x^2)^{n-1} = 2 \int_0^{\pi/2} \de \theta \, (\cos\theta)^{2n-1} = \sqrt{\pi} \, \frac{\Gamma(n)}{\Gamma(n+1/2)} \, ,
\end{displaymath}
where $\Gamma(n+1/2)=2^{-n} (2n-1)!! \, \sqrt{\pi}$. By using the fact that
\begin{displaymath}
(2n-1)!!=\frac{(2n)!}{(2n)!!} = \dfrac{(2n)!}{2^n  n!} \, ,
\end{displaymath}
one then finds
\begin{equation} \label{eq:up}
\Upsilon_{2n}(L) = 2 \zeta^{2n} \, \frac{(1-2n\alpha)^{2n-1}}{(2n)!} \, .
\end{equation}
From Eqs.~(\ref{eq:PLm-instantons}), (\ref{eq:ZL-instantons}) and (\ref{eq:up}), one ends up with
\begin{equation}
\begin{split}
\int_{-1}^{+1} \de \phi \, P_L (\phi) &= \frac{1}{Z_L (\zeta,\alpha)} \left[ \int_{-1}^{+1} d \phi \, \left( \delta(\phi-1) + \delta(\phi+1) \right) + \sum_{n=1}^{1/(2\alpha)} 2 \zeta^{2n} \, 
\frac{(1-2n\alpha)^{2n-1}}{(2n)!} \right] \\
& = \frac{1}{Z_L (\zeta,\alpha)} \sum_{n=0}^{1/(2\alpha)} 2 \zeta^{2n} \, 
\frac{(1-2n\alpha)^{2n-1}}{(2n)!} = 1 \, .
\end{split}
\end{equation}
\end{widetext}
The expression of $P_L (\phi)$ for the one-dimensional Ising model~\cite{antal} can be recovered as a particular case of
Eq.~(\ref{eq:PLm-instantons}) in the limit $\sigma \to 0$ ({\it i.e.}, for infinitely sharp domain walls) and for $\tilde{S}^\star = 2 J$.
In particular one then has
\begin{displaymath}
Z_L (\zeta,\alpha=0) = 2 \sum_{n=0}^{\infty} \, \frac{\zeta^{2n}}{(2n)!} = 2 \cosh \zeta \, ,
\end{displaymath}
and 
\begin{displaymath}
\begin{split}
P_L (\phi) &= \frac{1}{2 \cosh \zeta} \bigg[ (\delta(\phi-1) + \delta(\phi+1)) \\
& \qquad \qquad + \sum_{n=1}^{\infty} 2 (\zeta/2)^{2n} \frac{(1-\phi^2)^{n-1}}{n!(n-1)!} \bigg] \, .
\end{split}
\end{displaymath}
These expressions coincide with the results of Ref.~[\onlinecite{antal}].

\section{Instanton calculation for the one-dimensional Ising model}
\label{app:instantons_ising}

As stated in the main text, we consider the one-dimensional Ising model with periodic boundary condition which is described by the Hamiltonian
\beq
H[\{\sigma_i\}]=- J\sum_{i=1}^L\s_i\s_{i+1}-h\sum_{i=1}^L\s_i
\eeq
where $\s_{L+1}\equiv\s_1$, the lattice spacing which is as unity, and, contrary to the case studied in section \ref{sec:U_L}, only the thermodynamic limit $L \to \infty$ is considered.

We summarize the main (known) results about the model. The partition function can be computed using the transfer matrix method.\cite{baxter07} The transfer matrix is given by
\beq
\mathbf V=\begin{pmatrix}
e^{\tilde h+\tilde J} & e^{-\tilde J}\\
e^{-\tilde J} & e^{-\tilde h+\tilde J}
\end{pmatrix}
\eeq
where $\tilde h=\b  h$ and $\tilde J=\b J$ and the partition function is obtained as $Z_N=\mathrm{Tr}\mathbf V^N$.
One can easily diagonalize the transfer matrix with the rotation
\beq
\mathbf U=\begin{pmatrix}
\cos \theta & -\sin \theta\\
\sin \theta & \cos \theta
\end{pmatrix}
\eeq
where $1/\tan 2\theta=e^{2\tilde J}\sinh \tilde h$. The eigenvalues are given by
\beq
\l_\pm=e^{\tilde J} \cosh \tilde h \pm \sqrt{e^{2\tilde J}\cosh^2 \tilde h-2\sinh 2\tilde J} \:.
\eeq
The average magnetization $m=<\s_i>$ is then
\beq\label{mvsB}
m=\frac{e^{\tilde J}\sinh \tilde h}{\sqrt{e^{2\tilde J}\cosh^2\tilde h-2\sinh 2\tilde J}}
\eeq
and the two-point connected correlation function
\beq
G_c^{(2)}(|i-j|)=\langle \s_i\s_j\rangle-m^2=\sin^2(2\theta)\left(\frac{\l_-}{\l_+}\right)^{|i-j|}\,,
\eeq
which can be rewritten as
\beq
G_c^{(2)}(r)=\sin^2(2\theta)e^{-r/\xi}
\eeq
where $\xi = \left[\ln\left(\frac{\l_-}{\l_+}\right) \right]^{-1}$.
These expressions provide the magnetization and the two-point function at fixed external magnetic field $h$. However we would like to have the magnetization instead of the external field as the primary variable since we want to work with the effective action defined by the Legendre transform.(\ref{Legendre})

One can thus invert the relation in Eq. (\ref{mvsB}) to obtain
\beq
\frac{\l_+}{\l_-}=\frac{\sqrt{1-m^2(1-e^{-4\tilde J})}+e^{-2\tilde J}}{\sqrt{1-m^2(1-e^{-4\tilde J})}-e^{-2\tilde J}}\,,
\eeq 
so that, in the limit of very low temperature, one gets
\beq
\xi(m)^{-1}\simeq \frac{2e^{-2\tilde J}}{\sqrt{1-m^2}}
\eeq
and
\beq
G_c^{(2)}( r)=(1-m^2)e^{-r/\xi(m)}\:.
\eeq
These results are used in section \ref{sec:running_instantons}.

\section{Derivation of $\Gamma^{(3)}_k$ and $\Gamma^{(4)}_k$ from the one-dimensional Ising model}
\label{app:1PIvertices_ising}

We start from the calculation of the $3$-point correlation function in the Ising model. We need to compute
\beq
\langle \s_i \s_{i+r_1}\s_{i+r_1+r_2}\rangle \,.
\eeq
It is given by
\beq
\langle \s_i \s_{i+r_1}\s_{i+r_1+r_2}\rangle=\frac{1}{\l_+^N+\l_-^N}\mathrm{Tr}\left[\mathbf{\mathcal S} \mathbf V^{r_1}\mathbf{\mathcal S} \mathbf V^{r_2}\mathbf{\mathcal S} \mathbf V^{N-r_1-r_2}\right]
\eeq
where
\beq
\mathbf{\mathcal S}=\begin{pmatrix}
1 & 0\\
0 & -1
\end{pmatrix}\;.
\eeq
After diagonalizing the transfer matrix (see Appendix~\ref{app:instantons_ising}) and using that
\beq
\mathbf U^{-1}\mathbf{\mathcal S} \mathbf U=\begin{pmatrix}
\cos 2\theta & -\sin 2\theta\\
-\sin 2\theta & -\cos 2\theta
\end{pmatrix}\;,
\eeq
one obtains in the thermodynamic limit
\beq
\begin{split}
\langle \s_i \s_{i+r_1}&\s_{i+r_1+r_2}\rangle=c^3+cs^2\\
&\times\left(\left(\frac{\l_-}{\l_+}\right)^{r_1}+\left(\frac{\l_-}{\l_+}\right)^{r_2}-\left(\frac{\l_-}{\l_+}\right)^{r_1+r_2}\right)
\end{split}
\eeq
where $c=\cos 2\theta=m$ and $s=\sin 2\theta=1-m^2$ ($c$ should not be confused with the notation also used for the prefactor of the derivative term in the bare action).
The connected $3$-point correlation function is then given by
\beq
\begin{split}
\langle \s_i \s_{i+r_1}\s_{i+r_1+r_2}\rangle_c&=\langle (\s_i-\langle \s_i\rangle) (\s_{i+r_1}-\langle \s_{i+r_1}\rangle)\\
&\times(\s_{i+r_1+r_2}-\langle \s_{i+r_1+r_2}\rangle)\rangle\\
&=-2cs^2\left(\frac{\l_-}{\l_+}\right)^{r_1+r_2} \;.
\end{split}
\eeq
If we call $x_1=i$, $x_2=i+r_1$, $x_3=i+r_1+r_2$, the above result translates into
\beq
W^{(3)}(x_1<x_2<x_3)=-2c s^2\left(\frac{\l_-}{\l_+}\right)^{x_3-x_1}\;,
\eeq
so that the 3-point connected correlation function for generic arguments can be written as
\begin{widetext}
\beq
\begin{split}
&W^{3}(x_1,x_2,x_3)=-2c s^2\left[\left(\frac{\l_-}{\l_+}\right)^{x_3-x_1}\Theta(x_3-x_2)\Th(x_2-x_1)+\left(\frac{\l_-}{\l_+}\right)^{x_1-x_3}\Theta(x_1-x_2)\Th(x_2-x_3)\right.\\
&+\left(\frac{\l_-}{\l_+}\right)^{x_2-x_1}\Theta(x_2-x_3)\Th(x_3-x_1)+\left(\frac{\l_-}{\l_+}\right)^{x_1-x_2}\Theta(x_1-x_3)\Th(x_3-x_2)\\
&\left.+\left(\frac{\l_-}{\l_+}\right)^{x_2-x_3}\Theta(x_2-x_1)\Th(x_1-x_3)+\left(\frac{\l_-}{\l_+}\right)^{x_3-x_2}\Theta(x_3-x_1)\Th(x_1-x_2)\right]\,,
\end{split}
\eeq
where $\Theta(x)$ is the Heaviside step function. Neglecting the underlying lattice and performing the Fourier transform lead to
\beq
\begin{split}
W^{(3)}(p_1,p_2,p_3)=(2\pi)\d(p_1+p_2+p_3)\frac{4 c s^2 \xi^{-2}\left(\left(p_1p_2+p_1p_3+p_2p_3\right)-3\xi^{-2}\right)}{\left(p_1^2+\xi^{-2}\right)\left(p_2^2+\xi^{-2}\right)\left(p_3^2+\xi^{-2}\right)}\,.
\end{split}
\eeq
We now use the mapping between the Ising model and the $\varphi^4$ theory at low temperature and the relation between the connected $3$-point correlation function and the 1PI $3$-point vertex. \cite{Zinn} We finally obtain
\beq
\begin{split}
\label{eq:app_gamma3}
\G^{(3)}_k(p_1,p_2,p_3)&= -\tilde \G^{(2)}_k(p_1)\tilde \G^{(2)}_k(p_2)\tilde \G^{(2)}_k(p_3)W^{(3)}(p_1,p_2,p_3)\\
&=(2\pi)\d(p_1+p_2+p_3) \frac{c s^2}{2}\xi(\phi) \\&
\times \left[3\xi^{-2}(\phi)-\left(p_1p_2+p_1p_3+p_2p_3\right)\right]
\end{split}
\eeq
\end{widetext}
where $c=\phi$ and $s^2=\phi_{0,k}^2 - \phi^2$. Note that it is $\tilde \G^{(2)}_k( p)\equiv G_{c,k}^{(2)}( p)^{-1}$, obtained from $\tilde \G_k[\phi]=\G_k[\phi]+\frac{\beta}{2}\int \de q R_k(q)\phi(q)\phi(-q)$, which appears in Eq. (\ref{eq:app_gamma3}) and not $\G^{(2)}_k( p)$.
The calculation of $\Gamma_k^{(4)}$ can be done in an analogous way and leads to Eq. (\ref{Gamma4}). Although a cumbersome derivation, the higher orders can also be obtained along the same lines.


\begin{thebibliography}{99}
\bibitem{width}
Note that at small but finite temperature, the two $\delta$-functions in $\phi = \pm 1$
acquire a finite width due to the small gaussian thermal fluctuations of the fields around $\phi(x) = \pm 1$. 
Locally, the amplitude of such thermal fluctuations is simply related to $V^{\prime \prime} (\phi = \pm 1) = 2 r$.
Therefore the curvature of the two minima of $U_L(\phi)$ in $\phi = \pm 1$ is $2 \beta r / L$.
\bibitem{chi}
Note that in the $T \to 0$ limit, the correlation function has a simple exponential form, 
$G_c^{(2)}( r) \simeq e^{-r/\xi}$. As a consequence, the magnetic susceptibility becomes
$\chi = 2 \int  G_c^{(2)}( r)\, \propto \xi$ and $\kappa_{\infty} = 1/(2 \xi)$. This result can be
obtained analytically, as shown in Appendix~\ref{app:instantons_ising}.
\bibitem{footnote_contour}
Since the integrand in Eqs.~(\ref{eq:PLm-saddle}) and (\ref{eq:ULm-saddle}) is an analytic function, we can modify the contour in the complex plane.
\bibitem{footnote_dupuis}
The NPRG can also be formulated by starting with a microscopic theory defined on a lattice: see Ref. [\onlinecite{machado-dupuis}].
\bibitem{footnote_UV}
Strictly speaking, this is true if one chooses the cutoff function so that it diverges when $k \to \Lambda$.
\bibitem{footnote_spin_waves}
The return to convexity of the running effective potential has also been studied in the $O(N)$ model with $N>2$. In this case, the excitations above the ground state take the form of spin waves: see Refs. [\onlinecite{ringwald90},\onlinecite{tetradis92}].
\end{thebibliography}
\end{document}